\documentclass[aps,prb,twocolumn,unsortedaddress,superscriptaddress,showpacs]{revtex4-1}
\usepackage{color}
\usepackage{graphicx}
\usepackage{amsmath,amsfonts,amsthm} 

\begin{document}


\title{The crucial role of atomic corrugation on the flat bands and energy gaps of twisted bilayer graphene at the ``magic angle'' $\theta\sim 1.08^\circ$ }
\author{Procolo Lucignano}
\email{procolo.lucignano@spin.cnr.it}
\affiliation{CNR-SPIN, c/o Complesso di Monte S. Angelo, via Cinthia - 80126 - Napoli, Italy}
\author{Dario Alf\`e}
\affiliation{
Dept. of Earth Sciences  and 
London Centre for Nanotechnology 
University College London, 
Gower Street, London, WC1E 6BT, UK}
\affiliation{Universit\`a degli Studi di Napoli ``Federico II'',
Dipartimento di Fisica "Ettore Pancini", Complesso di Monte S. Angelo, via Cinthia - 80126 - Napoli, Italy}
\author{Vittorio Cataudella}
\author{Domenico Ninno}
\affiliation{Universit\`a degli Studi di Napoli ``Federico II'',
Dipartimento di Fisica "Ettore Pancini", Complesso di Monte S. Angelo, via Cinthia - 80126 - Napoli, Italy}
\affiliation{CNR-SPIN, c/o Complesso di Monte S. Angelo, via Cinthia - 80126 - Napoli, Italy}
\author{Giovanni Cantele}
\email{giovanni.cantele@spin.cnr.it}
\affiliation{CNR-SPIN, c/o Complesso di Monte S. Angelo, via Cinthia - 80126 - Napoli, Italy}




\date{\today}

\begin{abstract}
We combine state-of-the-art large-scale first principles calculations with a low-energy continuum model to describe the nearly flat bands of twisted bilayer graphene at the first magic angle $\theta=$1.08$^\circ$. 
   We show that the energy width of the flat  band manifold, as well as the energy gap separating it  from the valence and conduction bands, can be obtained only if 
   the out-of-plane relaxations are fully taken into account. The results agree both qualitatively and quantitatively  with recent experimental outcomes.
\end{abstract}


\maketitle

\section{Introduction}
The Moir\'e patterns originating from the commensurate rotation of
two graphene layers  with respect to each other have revealed, at small twist angles $\theta$, that the
Dirac cone picture breaks down~\cite{Cao:2018kn,Cao:2018ff,Codecido:2019wy}. Twisted Bilayer Graphene (TBG) at the "magic angle" $\theta\sim 1.08^\circ$  shows almost flat bands (FBs) at the Fermi energy, with a measured bandwidth as small as $\sim 10$ meV. The FBs manifold can host up to four electrons above the Fermi energy and four holes below it and is  separated by an energy gap of $\sim 50$  meV from both higher and lower energy bands. When an external gate tunes the system chemical potential within these gaps, a clear band insulating phase appears. 	
A second, unexpected, insulating phase shows up at half-filling of the FB manifold, both on the electron and on the hole side ($\pm 2$ electrons with respect to charge neutrality). 
After electrostatic doping, achieved by gating the structure, unconventional superconductivity, with a 1.7 K critical temperature, appears in a strong pairing regime, with a phase diagram very similar to that of the underdoped cuprates.
The latter two features are attributed to enhanced electron-electron or electron-phonon interaction within the FBs, respectively,  and are currently under study~\cite{Choi:2018cu,Sboychakov:2018tp}. 

This remarkable scenario reveals how the twist angle can be used as a further degree of freedom~\cite{RibeiroPalau:2018ki} to combine two-dimensional (2D) materials exhibiting vertical stacking, to implement desired 
properties~\cite{Geim:2014hf, PhysRevMaterials.1.014002,Borriello:2012ja,Cantele:2009de}.
The twisted lattice geometry gives rise to topological
properties of TBG~\cite{Song:2018ul,Hejazi:2019dz,Liu:2018}, by contrast to  
conventional topological materials~\cite{RevModPhys.82.3045}, where topological properties are mostly due to  spin-orbit interactions~\cite{0034-4885-78-10-106001,PhysRevB.78.035336} and Brillouin zone topology.


In this paper we  focus on the   band insulating phase, that requires  an accurate description of the single-electron properties  determining the band structure of the TBG. 
These electronic properties have been addressed mostly adopting  continuum effective models, focusing only on low-energy states~\cite{LopesdosSantos:2007fg,LopesdosSantos:2007fga,Mele:2010jg,Bistritzer:2011ho,Moon:2013bc,Koshino:2018hm}, tight binding~\cite{Jung:2013ix,Sboychakov:2015jx,Fang:2016iq,Lin:2018fy,Kang:2018et,Angeli:2018wy} or calculations using time-dependent Schr\"odinger approaches~\cite{Bercioux:2018,PhysRevB.97.125136}. The point is that  the unit cell, at the first magic angle $\theta=1.08^\circ$, contains 11164 atoms, and this makes it impossible to perform a full many body calculation 
(details about the lattice geometry and the reciprocal space can be found in appendix \ref{sec:app-geom}). 
Moreover, even a full description of the system in the framework of ab-initio density functional theory (DFT), including atomic relaxations, remains challenging (the corresponding supercell is described by an hexagonal lattice with an in-plane lattice parameter of $\sim 120$ {\AA}).
To the date, there are very few first-principles calculations~\cite{Kang:2018et,Song:2018ul} on TBG at small twist angles.
For example, in Ref.~\onlinecite{Kang:2018et} the energy bands obtained from DFT calculations carried out on the unrelaxed structure, are reported. However, the ab initio  results at the magic angle $\theta\sim 1.08^\circ$ show no gap between the FB manifold and the closest lower band,  although a band set consistent with the experimental outcomes is reproduced at a  larger angle, $\theta=1.30^\circ$, corresponding to a smaller unit cell (7804 atoms to be contrasted with 11164 at the first magic angle).
However, since the experimental uncertainty over the measured angle in Ref.~\onlinecite{Cao:2018kn} is of the order of $\Delta \theta \sim 0.01^\circ-0.02^\circ$,  a thorough  theoretical description at $1.08^\circ$ is demanding.

In this manuscript we present a fully ab initio DFT calculation of the electronic structure of TBG at $\theta=1.08^\circ$   showing that   a band structure consistent with that measured in Ref.~\onlinecite{Cao:2018kn} is obtained provided that the out-of-plane atomic relaxations are fully taken into account. 
In particular, the  occurrence of energy gaps between the FBs and the lower and higher energy bands emerge as a direct consequence of the corrugation due to the out-of-plane displacements.

Such result is strictly related with the nature of the FBs close to the Fermi energy: indeed, we can infer that the appearance of these bands is the result of the inter-layer vdW interaction, whose effect can be tuned by controlling the twist angle.
The out-of-plane relaxation of the atomic positions obviously modifies the strength of the interaction, as we are going to discuss in the following.

The paper is organized as follows. In sec. II we describe the ab-initio calculation with particular attention to the geometric optimization of the superlattice. 
In Sec. III we introduce the  low energy effective continuum model of Ref.s~ \onlinecite{LopesdosSantos:2007fg,LopesdosSantos:2007fga,Mele:2010jg,Bistritzer:2011ho,Moon:2013bc,Koshino:2018hm} and specialize it to describe our optimized structures. In Sec. IV we show our numerical results. In Sec. V we summarize our findings. Appendices A, B and C describe details of the geometrical structure, of the DFT ab initio calculation and of the low energy continuum model, respectively.
  
\begin{figure}[t!]
\includegraphics*[scale=0.32]{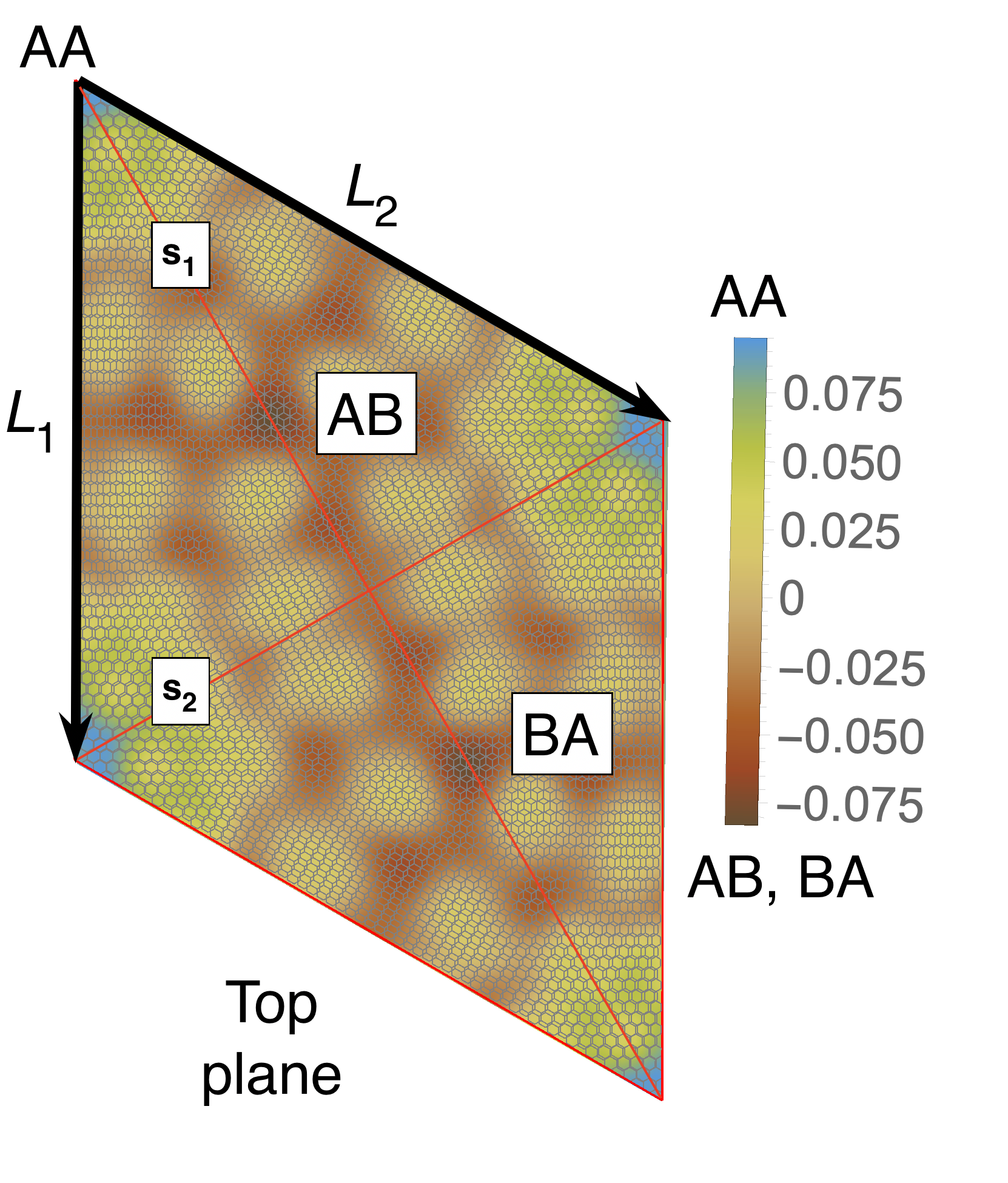}
\caption{\label{fig:structure} Color map of the top layer relaxation (color labels are expressed in {\AA} and correspond to the difference $\Delta z = z - z_{avg}$ between the actual $z$ coordinate and the average coordinate $z_{avg}$ in the top plane) as obtained from first principles DFT calculations.  AA and AB/BA stacking regions
are highlighted, as well as the supercell sides $\mathbf{L}_1,\mathbf{L}_2$. }
\end{figure}
 
\section{Geometric Optimization}
Although previous studies have pointed out that the atomic corrugation due to inter-plane vdW interaction might have relevant effects on the band structure and on the effective point symmetries \cite{Uchida:2014,Fasolino:2015,Tomanek:2018,Angeli:2018wy}, they were mainly based on molecular dynamics and classical interatomic potentials~\cite{Fang:2016iq,Gargiulo:2018bj,Jain_2016}, which, as such, can only give a partial answer to the problem posed.

\begin{figure*}[t!]
	\includegraphics*[height=3.7cm]{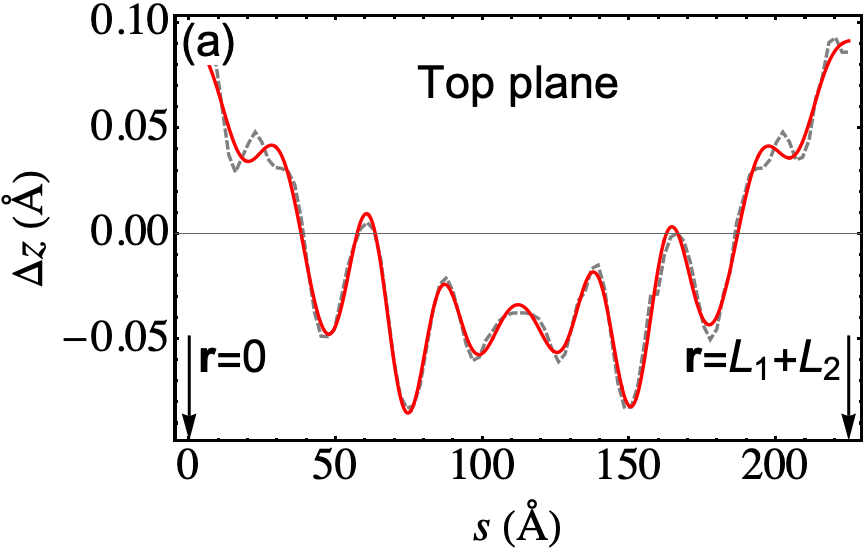}
	\includegraphics*[height=3.7cm]{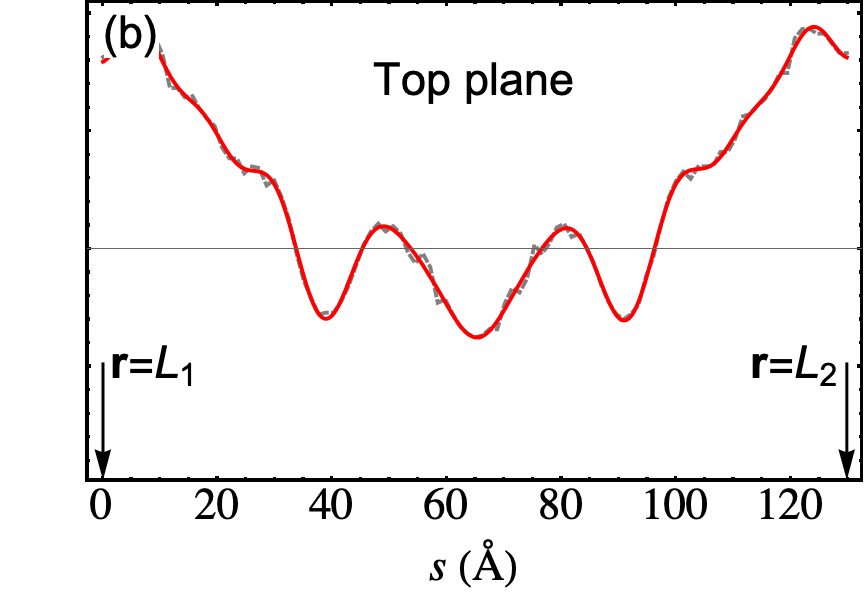}\hspace*{0.2cm}
	\includegraphics*[height=3.7cm]{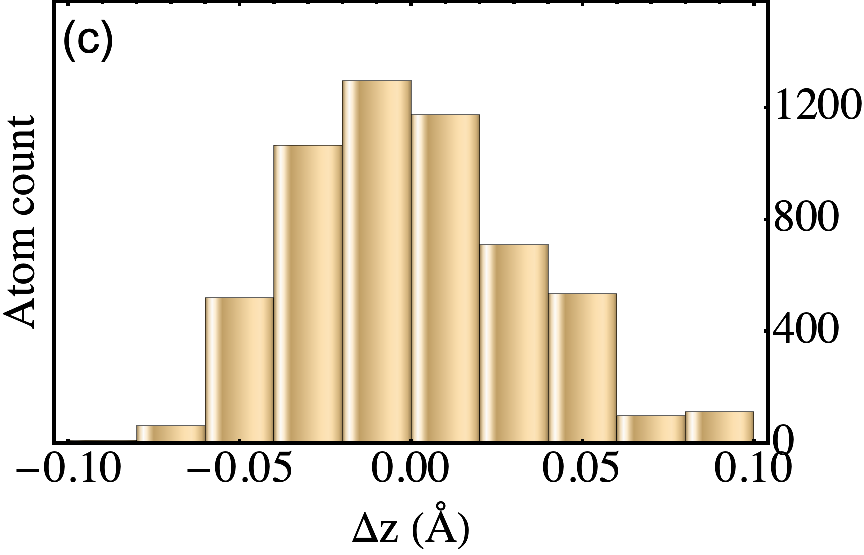}\vspace*{0.2cm}\\
	\includegraphics*[height=3.7cm]{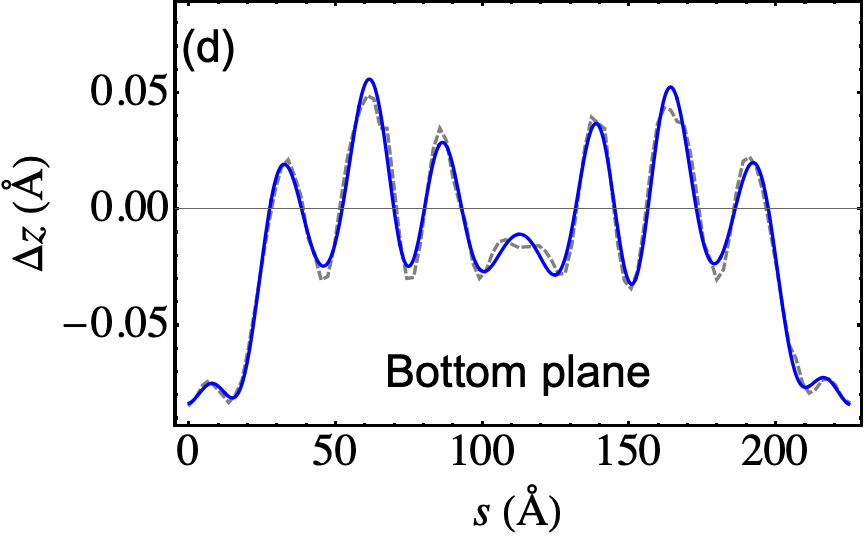}
	\includegraphics*[height=3.7cm]{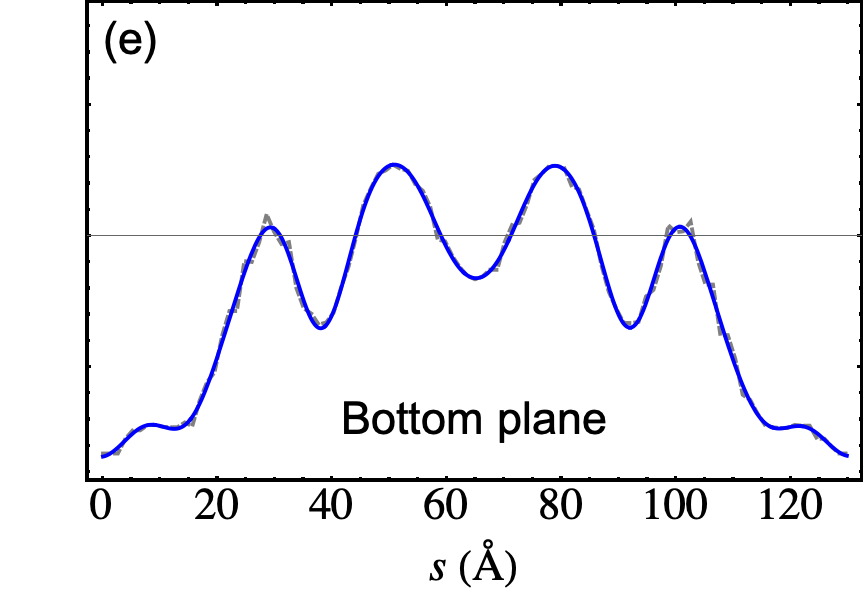}\hspace*{0.2cm}
        \includegraphics*[height=3.7cm]{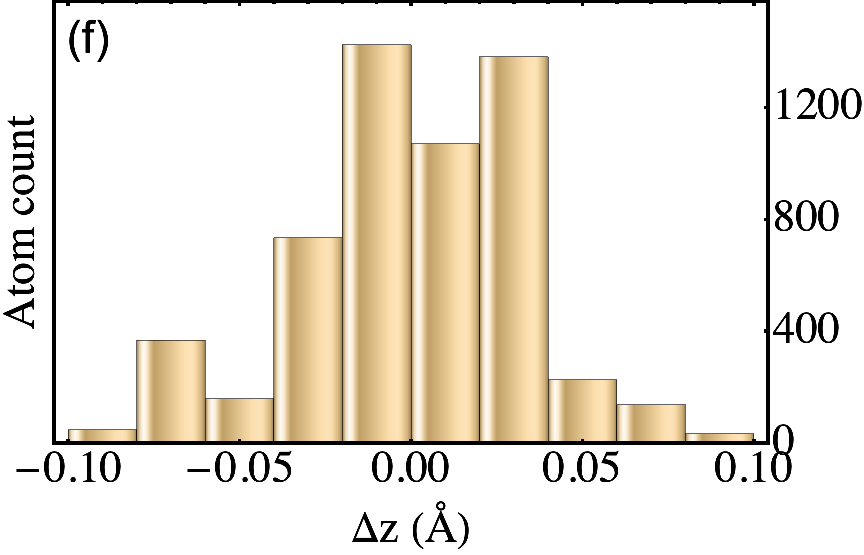}
	\caption{\label{fig:deformazioni} Analysis of the atomic corrugation. In the top panels we illustrate the characteristics of the top plane: (a) corrugation profile along the line $\mathbf{s}_1=s\left(\mathbf{L}_1+\mathbf{L}_2\right),0\le s \le 1$ shown in Fig.~\ref{fig:structure}. $\Delta z = z - z_{avg}$ is the deviation of the $z$ coordinate with respect to the average $z$ of the top plane, $z_{avg}$. Thin dashed lines correspond to real atomic $z$-coordinate, while full colored lines are interpolating functions.
(b) Corrugation profile along the line $\mathbf{s}_2=\mathbf{L}_1 + s \left(\mathbf{L}_2-\mathbf{L}_1\right), 0 \le s \le 1$ shown in Fig.~\ref{fig:structure}.
(c) The histogram represents the atomic population binned according to the $\Delta z$.
The same features are shown in panels (d), (e) and (f), respectively, for the the bottom plane, with $\Delta z$ built according to the average $z$ coordinate of the bottom plane.}
\end{figure*}

Here, DFT calculations, using the vdW-DF2 exchange-correlation functional \cite{Hamada}, have been carried out using the Vienna Ab initio Simulation Package (VASP) \cite{Kresse1}. The atomic positions have
been fully optimized, as detailed in appendix \ref{sec:app-DFT}.
The calculation required from 2880  (distributed over 80 nodes) up to  5760 physical cores (distributed over 160 nodes)  of a Cray XC-40 machine, over a period of about 30 days.

The outcome of the geometry optimization is depicted in Fig.~\ref{fig:structure}, where we use a color map to show the out of plane deformation of the top layer. Inspection shows that atomic relaxation tends to increase the interplane distance in correspondence to the AA stacking regions, and to decrease it in the AB regions.  Out-of plane displacements are modulated on length scale, intermediate between  the Moir\'e and the graphene periodicity, and seem to be in agreement with the emergent D6 symmetry described in Ref.~\onlinecite{Angeli:2018wy}. As we will show in the following, these geometric properties have severe consequences on the electronic band structure of the system. 
Details on the optimized structure are shown in Fig.~\ref{fig:deformazioni}. In the top panels we illustrate the characteristics of the TBG top plane: in particular, in Fig.s~\ref{fig:deformazioni}(a)-(b) we show the corrugation profile respectively along the line $\mathbf{s}_1=s\left(\mathbf{L}_1+\mathbf{L}_2\right),0\le s \le 1$ and the line $\mathbf{s}_2=\mathbf{L}_1 + s \left(\mathbf{L}_2-\mathbf{L}_1\right), 0 \le s \le 1$ (both shown in Fig.~\ref{fig:structure}). In Fig.~\ref{fig:deformazioni}(c) we show an histogram representing the atomic population binned according to the $z$-coordinate of the atoms. 
The same features are shown in Fig.s~\ref{fig:deformazioni}(d)-(e)-(f) for the the TBG bottom plane. The corrugations show clear oscillations, along the lines $\mathbf{s}_1$ and $\mathbf{s}_2$, however a simple analytical expression interpolating between the atomic position is not easily accessible, due to large harmonic content of the oscillatory behavior. The peaks of the histograms define an average $z$-coordinate $z_{avg}^{top}$ for the top and $z_{avg}^{bottom}$ for the bottom plane. Their difference defines the average interlayer spacing $z_{avg}^{top}-z_{avg}^{bottom}=3.408$ {\AA}  that is half way between the equilibrium distances of 3.31 {\AA} and 3.496 {\AA} between consecutive planes in graphite with AB Bernal and AA stacking, respectively (both calculated using the same vdW-DF2 exchange-correlation functional). Within each plane, the atomic displacements occur within an interval of about 0.2 {\AA} ($\pm$0.1 {\AA} with respect to the average $z$ in each plane).
\section{Effective Continuum model}
For increasing the understanding of the ab initio results,
we also describe a  continuum model generalizing the model proposed in Ref.s~\onlinecite{LopesdosSantos:2007fg,Bistritzer:2011ho,Moon:2013bc,Koshino:2018hm}, providing an effective low-energy band structure  which shows a remarkable agreement with the DFT calculation. Our results can be viewed as an accurate single-particle description of TBG at the first magic angle and used as starting point for a full many-body calculation taking into account electronic correlations.
In the following we shortly summarize the model, referring to Refs.~\onlinecite{LopesdosSantos:2007fga,Moon:2013bc,Koshino:2018hm} for further details. At small twist angles, the Moir\'e period $L_M$ is much longer than the lattice constant $a$.
The superlattice Mini Brillouin Zone (MBZ) extends over a tiny small area of the graphene BZ, and it is an hexagon whose vertices are the two Dirac points $K^{(1)}_{\xi}$ and $K^{(2))}_{\xi}$ of the two layers after rotation (compare small and large hexagons in Fig.~\ref{fig:BZ}), where $\xi=\pm 1$ is the valley index. Close to these points the single-layer graphene spectrum can be safely assumed to be linear and a low energy (long wavelength) Hamiltonian of each layer $l=1,2$ can be used:

\begin{equation}
H_\xi^{(l)}(\mathbf{k}) = -\hbar v_F \left\{R\left[(-1)^{l+1}\theta/2\right](\mathbf{k}-\mathbf{K}_\xi)\right\} \cdot (\xi \sigma_x,\sigma_y)
\end{equation}
where $\hbar v_F/a = 2.1354$ eV and $\sigma_x,\sigma_y$ are Pauli's matrices.
In the following we neglect the intervalley mixing, because of the huge distance between the two graphene valleys on the MBZ scale. Hence each valley can be studied separately. 
In the presence of inter-layer interaction the bilayer system can be described by the matrix Hamiltonian 
\begin{equation}
H_\xi(\mathbf{k})=
\left(\begin{array}{cc}
H_\xi^{(1)}(\mathbf{k}) & U_\xi^\dagger\\
U_\xi & H^{(2)}_\xi(\mathbf{k})
\end{array}
\right).
\end{equation}
The off-diagonal coupling terms are expressed in terms of overlap integrals $u,u'$ (see appendix \ref{sec:app-ECM}). 
The parameters $u$ and $u'$ 
are calculated in Ref.~\onlinecite{Koshino:2018hm} at $\mathbf{k}=\mathbf{K}_\xi^{(l)}$ and kept constant when calculating the band structure for all the $k$ points in the MBZ. As the MBZ is a small hexagon of side $\left|\mathbf K_{\xi}^{(1)} -\mathbf K_{\xi}^{(2)}\right|\sim \left| \mathbf K_\xi^{(l)} \right| \theta$, it seems to be  a reasonable approximation, in particular at small twist angles. In our calculations, in order to give a minimal model capable of describing (at least) the low energy properties of the ab-initio band structure, we do not calculate $u,u'$ but use them as fitting parameters. In the following we will show that the fitted parameters are relatively close to (but quantitatively different from) those obtained performing the hopping integrals shown in the appendix \ref{sec:app-ECM}. 
Such discrepancy can be ascribed to the fact that a low energy effective model obtained expanding a tight-binding Hamiltonian around the $\mathbf{K}_\xi$ point does not entail all the complexity of the full-ab-initio approach, but nevertheless can constitute a relevant tool to get closer and closer to the desired solution, with an accurate choice of the model parameters.

\begin{figure*}[t!]
	\includegraphics*[scale=0.25]{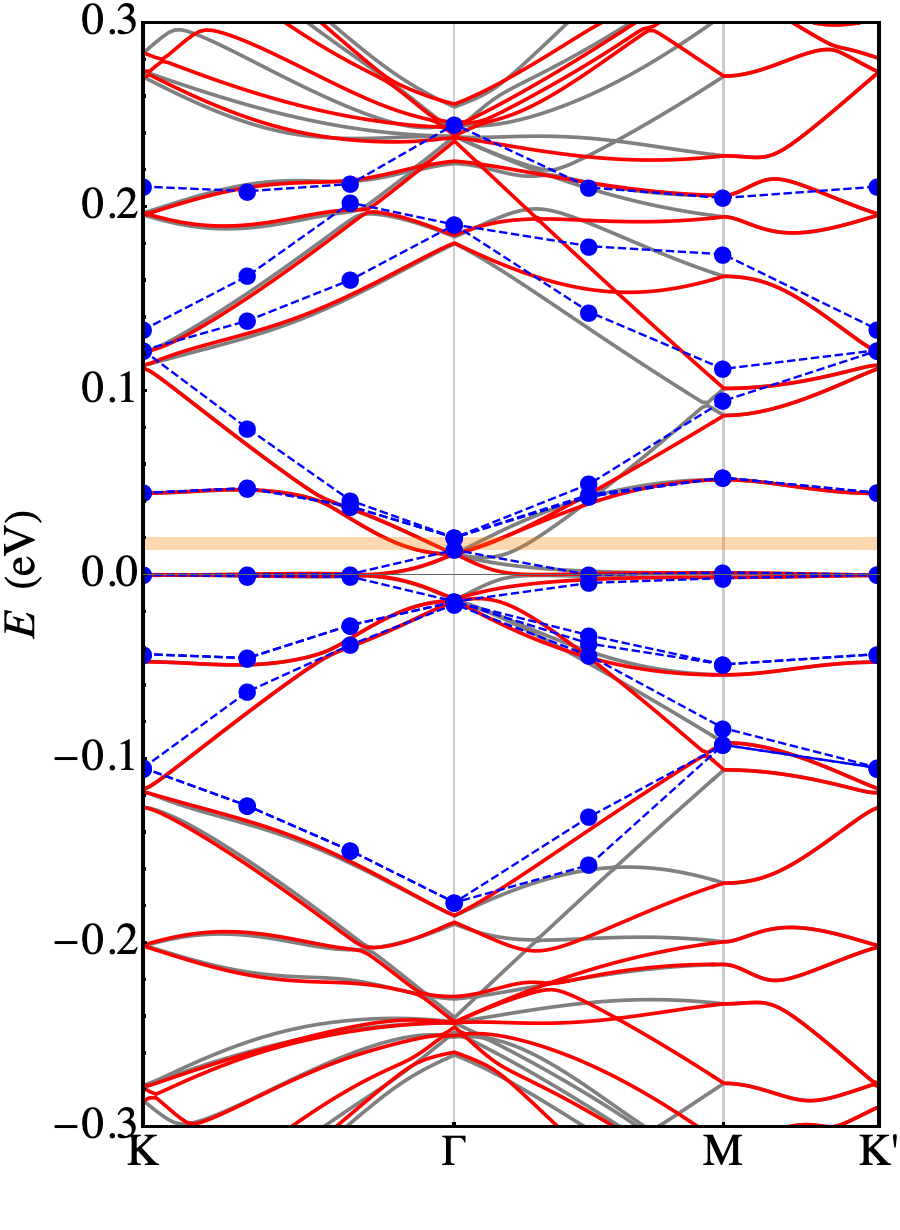}\hspace*{0.3cm}
	\includegraphics*[scale=0.25]{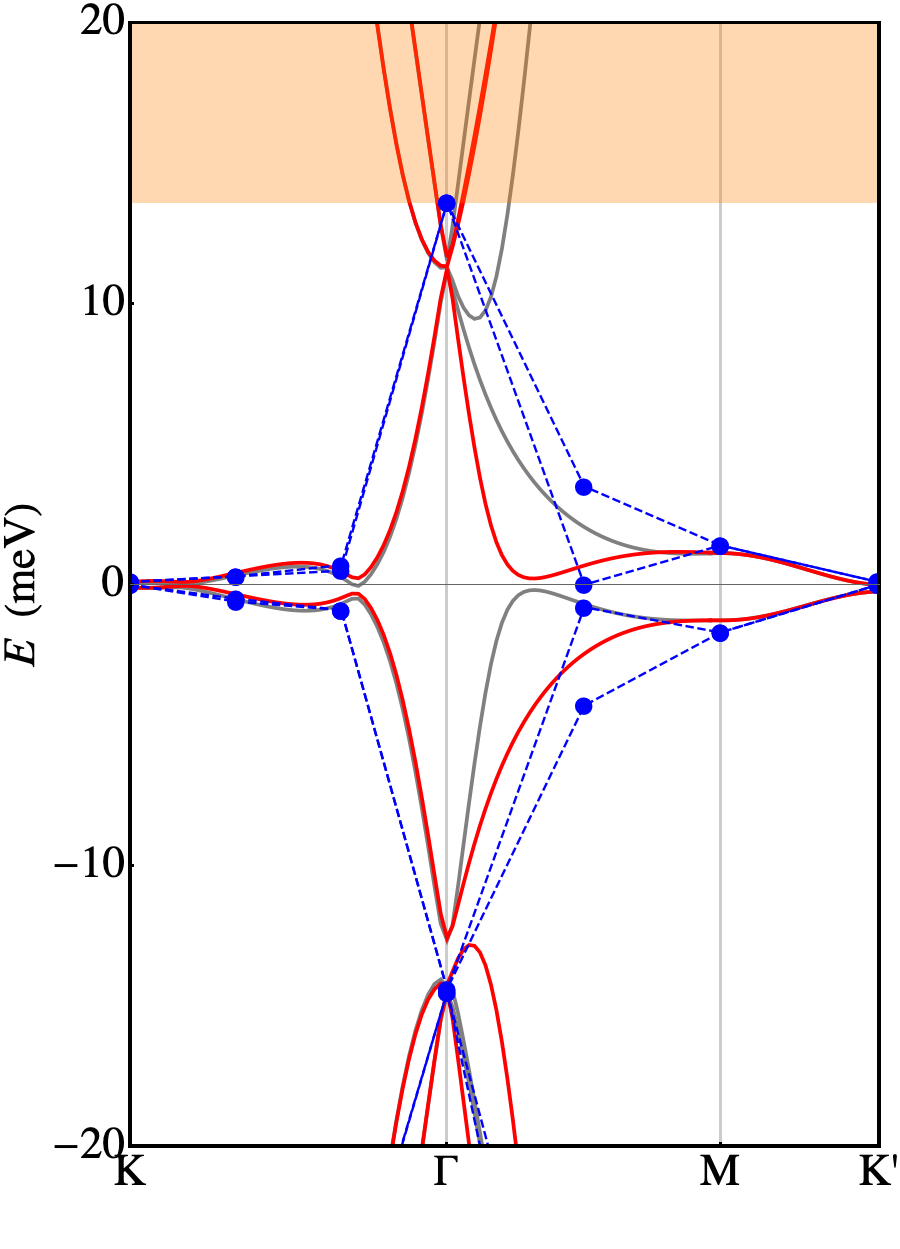}
	\caption{\label{fig:bands_unrelaxed}(Left) Unrelaxed band structure along the $K-\Gamma-M-K'$ line as derived from the continumm model (solid red and gray
		lines represent bands deriving from $\xi=1$ and $\xi=-1$ valleys, respectively). Blue dots are the results of the ab initio calculation.
		(Right) Zoom around the Fermi energy of the top panel, highlighting the nearly FB. Zero energy corresponds to the Fermi level. The orange-shadowed regions correspond to the energy ranges of the DFT gaps. Note the absence of any gap between the four flat bands and the low energy part of the spectrum.}
	\label{unrelaxed}
\end{figure*}

The wave function is calculated as a linear superposition of plane waves of momentum $\mathbf{G}$ where $\mathbf{G}$ are  reciprocal lattice vectors. The $\mathbf{G}$ point expansion extends, in principle, over the full (infinite) set of $\mathbf{G}$ vectors. However, for numerical purposes this set has to be truncated. We choose a cutoff radius $G_{cut}$, and keep only the $\mathbf{G}$ vectors inside the sphere of radius $G_{cut}$. It turns out that the number $N_G$ of required vectors to converge the lowest energy states is rather small. The low-energy continuum Hamiltonian matrix has the dimension $D=4 N_{G}$, and  $N_G=19$ (as in the example reported in the figure) allows for a good convergence in an energy shell of few hundreds on meV around the Fermi energy, whereas full convergence, i.e. band energies converged within less than 1 meV, is achieved with $N_G=37$.

\section{Results and Discussion}
\begin{figure*}[t!]
	\includegraphics*[scale=0.25]{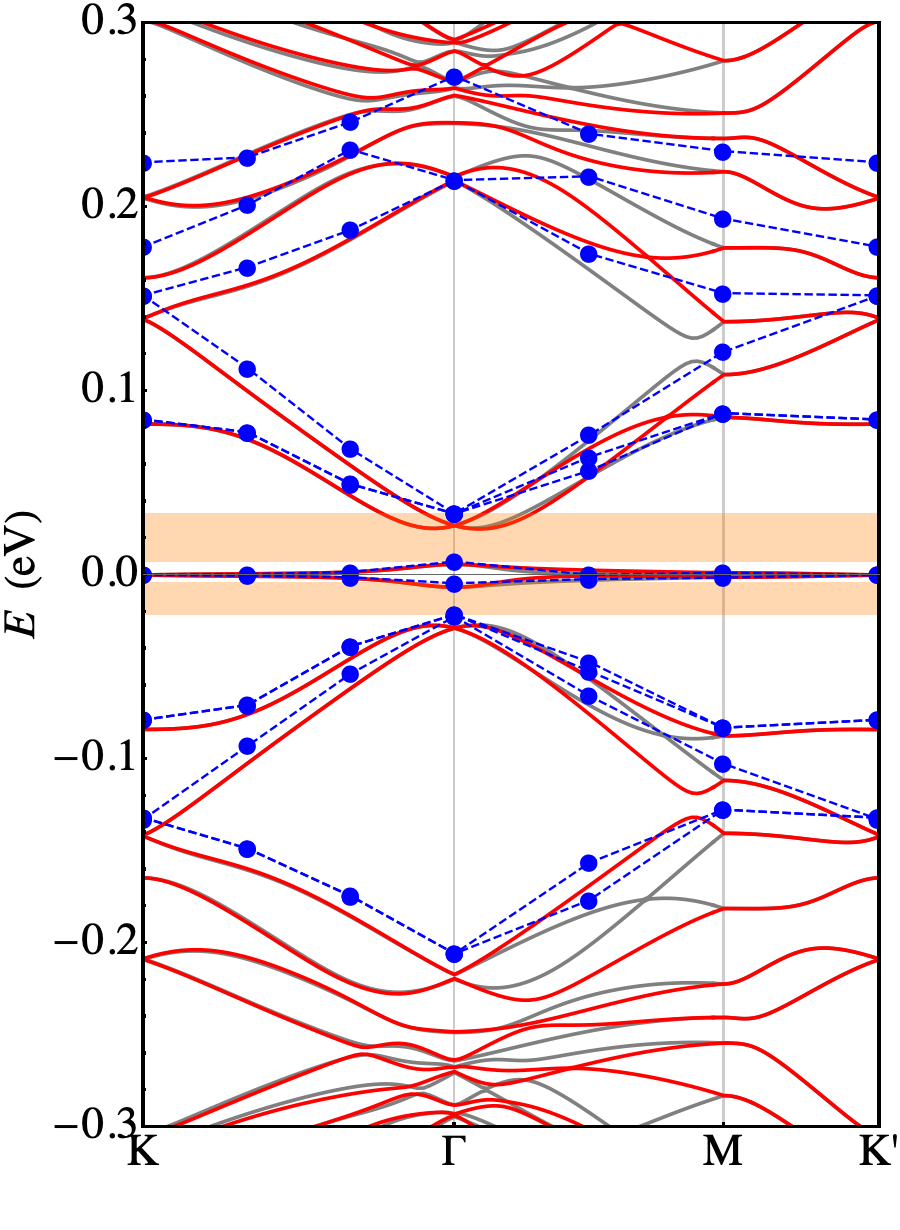}\hspace*{0.3cm}
	\includegraphics*[scale=0.25]{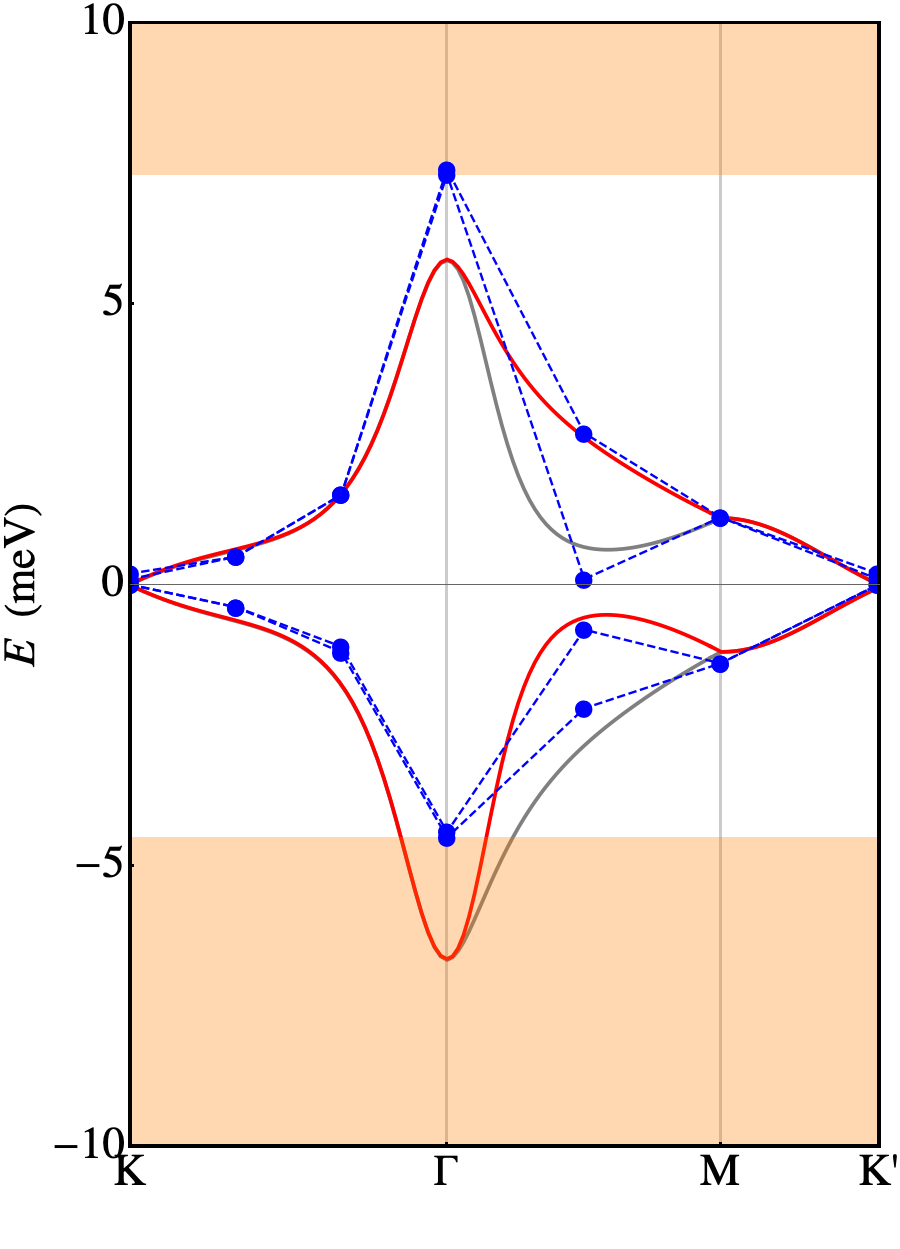}
	\caption{\label{fig:bands_relaxed}(Left) Relaxed band structure along the $K-\Gamma-M-K'$ line as derived from the continuum model (solid red and gray lines represent bands deriving from $\xi=1$ and $\xi=-1$ valleys, respectively). Blue dots are the results of the ab initio calculation.
		(Right) Zoom around the Fermi energy of the top panel, highlighting the nearly FB. Zero energy corresponds to the Fermi level. The orange-shadowed regions correspond to the energy ranges of the DFT gaps. Note the opening of the gap between the four flat bands and the low energy part of the spectrum, caused by the relaxation of the atomic coordinates.}
	\label{relaxed}
\end{figure*}

In Fig.~\ref{unrelaxed} we represent the band structure calculated in the absence of structural relaxation.
The blue dots are the outcome of a DFT ab-initio calculation,  while the full curves are obtained diagonalizing the continuum model  in the two valleys: $\xi=\pm 1$. The parameters  $u=u'=0.1085$ eV are obtained after the fitting procedure on the ab-initio points. The results show a reasonable agreement between these two approaches.
Here we discuss some relevant features emerging from our numerical calculations.

First of all we may notice that the FB has a dispersion of  $\sim$ 20 meV (calculated ab-initio) which is almost twice as the one measured in the experiment of Ref.~\onlinecite{Cao:2018ff}. Another relevant issue is that the unrelaxed ab-initio calculation is not able to reproduce the  gap between the FB and the first excited bands (both on the electron and on the hole side) that are responsible for the  band insulating phases. Our  calculation performed adopting the low-energy continuum model shows a good agreement with the ab-initio calculation and reproduces all its relevant features. Few discrepancies shows up, only when zooming in the very fine details of the FB (see the bottom panel of Fig.~\ref{unrelaxed}), that are not relevant as they do not change the way the data compare with experiments. 
However, these results point in the direction that it is not possible to interpret experimental data using an unrelaxed structure.

\begin{figure}
\includegraphics*[width=9cm]{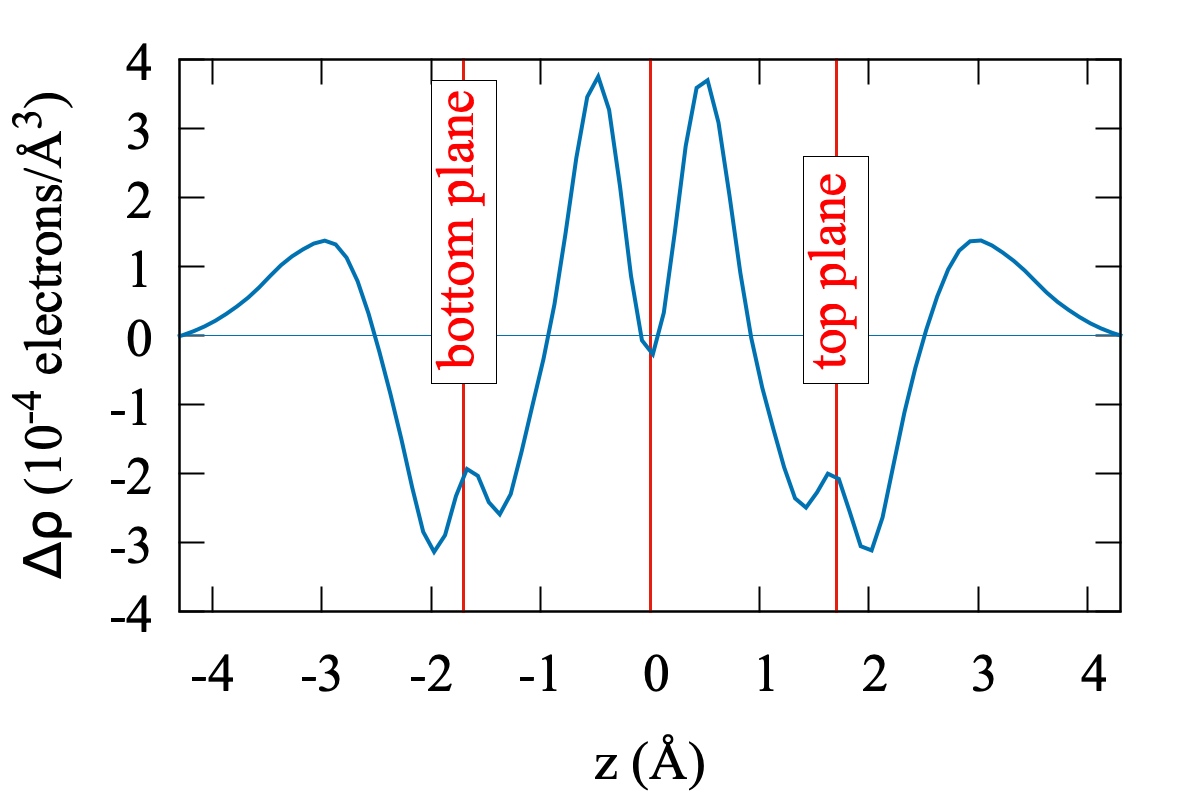}
\caption{\label{fig:CT} Computed DFT charge transfer: positive (negative) values correspond to filling (depletion) of
electron charge}
\end{figure}

Relaxation of the structure drastically changes the scenario, with a substantial agreement with the experiments. 
Plots of the relaxed band structures are shown in Fig.~\ref{relaxed}.
The parameters $u=0.0761$ eV, $u'=0.1031$ eV are again obtained fitting the ab-initio band structure.
The FB, now extends for $\sim 12$ meV around the Fermi level (calculations performed with larger supercells with $z$-axis = 12 and 14 {\AA} show that this number is subject to an error of approximately 3 meV). That bandwidth is in good agreement with the one  measured in experiments ($\le$ 10 meV, see Ref.~\onlinecite{Cao:2018kn}). It is separated by a gap of 26 meV (16 meV) from the highest occupied (lowest unoccupied) bands to be compared with the thermal activation gap of $\sim$ 40 meV measured in experiments. Such discrepancy is not much larger than the convergence error in the DFT calculations.
It is clear that the vdW inter-plane interactions, despite being weak, play a crucial role in determining the details of the TBG at the meV level. This is confirmed by the computed charge transfer, shown in Fig. \ref{fig:CT}. It is defined as $\Delta\rho(\mathbf{r}) =
\rho(\mathbf{r}) - \rho_1\mathbf{r}) - \rho_2(\mathbf{r})$ where $\rho(\mathbf{r})$ is the total charge density of TBG, whereas $\rho_i(\mathbf{r}), i=1,2$ are the charge densities of the top and bottom plane, respectively, calculated removing the other plane from the supercell at frozen atomic positions. Fig. \ref{fig:CT} shows the average charge transfer over planes orthogonal to the $z$ axis. It turns out that an electronic charge depletion shows up in correspondence of both planes and most of the charge is redistributed in the interplane region.

\section{Conclusions}
To summarize, our calculations can finally give a clear explanation of some of the most striking features of the electronic structure of TBG at the first magic angle $\theta=1.08^\circ$. In particular, the extension of the FBs and the presence of band gaps separating them from excited states,  both on the negative and positive energy side, can be explained successfully in terms of atomic out-of-plane
displacements. By allowing a full ab initio structural optimisation, a non-negligible atomic corrugation shows up in both the graphene layers.  As expected from simple electrostatic arguments, the interlayer distance gets larger (smaller) in correspondence of AA  (AB/BA) stacking regions, with a maximum
(minimum) distance of $\sim$3.68 {\AA} (3.28 {\AA}).  Such corrugation is a direct consequence of the interplay between vdW interaction and twisting of the graphene layers. It implies a decrease of the FB bandwidth of $\sim$  4 meV and induces gaps between the FB and the closer bands, in good agreement with the experimental findings.
Our ab initio results can also be interpreted in terms of a simple continuum model in which interplane hopping potentials have been used as fitting parameters. This simple model reproduces with reasonable accuracy the electronic structure and could pave the way for further investigations, to better describe also the other relevant phases of the TBG at small twist angle, i.e. the superconducting and the correlated insulating one.

\begin{acknowledgments}
We thank D. Bercioux,  M. Fabrizio and A. Tagliacozzo for fruitful discussions. We acknowledge use of the Monsoon2 system, a collaborative facility supplied under the Joint Weather and Climate Research Programme, a strategic partnership between the UK Met Office and the Natural Environment Research Council.
\end{acknowledgments}

G.C. and P.L. contributed equally to this work.

\renewcommand\thefigure{A\arabic{figure}}    
\setcounter{figure}{0}
\appendix
\section{Lattice Geometry and Reciprocal Space}
\label{sec:app-geom}
Let $\mathbf{a}_i$, $i=1,2$ be the vectors defining the graphene primitive cell, where $\mathbf{a}_1 = a(1,0)$, $\mathbf{a}_2 = a(1/2,\sqrt{3}/2)$ and $a\sim 0.246$ nm is the lattice constant. The corresponding reciprocal lattice vectors are $\mathbf{b}_1 = \left(2 \pi/a\right)(1,-1/\sqrt{3})$, and $\mathbf{b}_2 = \left(2 \pi/a\right)(0,2/\sqrt{3})$.
In the absence of geometric relaxation, let us consider, as starting point, the unrotated bilayer, with perfect AA stacking (each C atom in the first layer laying exactly on top of a C atom in the second layer). Choosing a pair of stacked C atoms, each belonging to one of the layers, the twisted bilayer at angle $\theta$ can be obtained by rotating the first and the second layer around the axis passing through these atoms (that indeed are fixed points of the rotation), by $-\theta/2$ and $\theta/2$ respectively. After rotation, the Bravais direct lattices of the first and of the second layer are described by the vectors $\mathbf{a}_i^{(l)}=R(\mp\theta/2)\mathbf{a}_i$ and the reciprocal lattice by the vectors $\mathbf{b}_i^{(l)}=R(\mp\theta/2)\mathbf{b}_i$ where $l=1,2$ identifies the layer and 
$R(\theta)$ is a two-dimensional matrix describing the rotation by $\theta$.
For an arbitrary rotation angle, the resulting structure shows a Moir\'e pattern but is aperiodic and cannot be described through a Bravais lattice because the periods of the two layers are, in general, incommensurate. However, periodic structures can be achieved when $\theta$ is the angle between two lattice vectors $\mathbf{d}_1=n \mathbf{a}_1 + m \mathbf{a}_2$ and $\mathbf{d}_2=m \mathbf{a}_1 + n \mathbf{a}_2$ with $(n,m)$ being an arbitrary pair of integers. The points at $\mathbf{d}_1$ and $\mathbf{d}_2$ merge after the rotation of the two planes and the lattice vectors of the Moir\'e supercell (MS) are thus given by $\mathbf{L}_1= n\mathbf{a}_1^{(1)} + m \mathbf{a}^{(1)}_2= m\mathbf{a}_1^{(2)} + n \mathbf{a}^{(2)}_2$  and $\mathbf{L}_2=R(\pi/3) \mathbf{L}_1$. The rotation angle can be expressed in terms of the integers $n,m$ as $2 \cos\theta = (m^2+n^2+4mn)/(m^2+n^2+mn)$.
The magic angle $\theta=1.08^\circ$ corresponds to $(n,m)=(31,30)$, with the number of atoms in the unit cell given by $ N = 4\; |(\mathbf{L}_1 \times \mathbf{L}_2) |/ | (\mathbf{a}_1 \times \mathbf{a}_2)| =11164$. As $n=m+1$, the lattice constant $L=|\mathbf{L}_1|=|\mathbf{L}_2|= a |m-n| /\left[2 \sin(\theta/2)\right]\sim12.78$ nm  is coincident in this case with  the  Moir\'e pattern period $L_M= a /\left[2 \sin(\theta/2)\right]$.~\cite{Moon:2013bc}

The top panel of Fig.~\ref{fig:structure} shows the atomic structure of the TBG at $\theta=1.08^\circ$ (four unit supercells are shown).  The Moir\'e patterns originating from regions with different stacking are highlighted: AA and AB/BA, AB (BA) corresponding to the stacking of a C atom in the top (bottom) layer and the center of an hexagon in the other layer.

\begin{figure}
	\includegraphics*[scale=0.3]{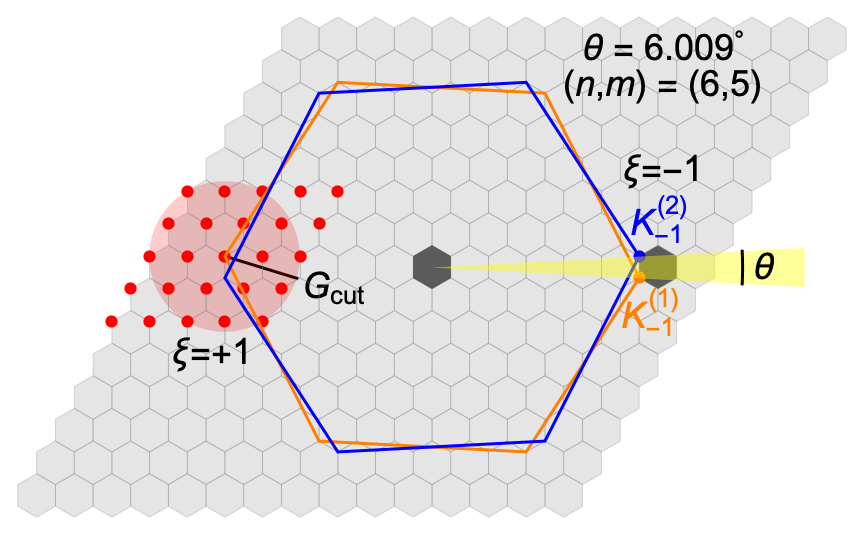}
	\caption{\label{fig:BZ}The reciprocal lattice of the Moir\'e supercell. The central Mini Brillouin Zone is highlighted with a dark hexagon and periodically replicated (light gray hexagons). The rotation angle is $\theta=6.009^\circ$ corresponding to $(n,m)=(6,5)$ (a larger angle than the first magic angle is considered to produce a more clear representation of the MBZ). $\xi=\pm1$ denotes the valley in the BZs of the single layers, and $K_\xi^{(l)}$ the $K$ point of layer $l=1,2$ at valley $\xi$. The bigger (blue and orange) hexagons represent the BZs of the layers after rotation and the rotation angle $\theta$ is highlighted. Red dots correspond to a uniform grid of $G$ vectors of the supercell reciprocal lattice. Only $G$ vectors up to a given distance from a $K$ point (red-shaded circle) are included in the expansion required in the continuum model (see text). }
\end{figure}

The reciprocal lattice vectors for the Moir\'e pattern are obtained as $\mathbf{G}_i = \mathbf{b}_i^{(1)}-\mathbf{b}_i^{(2)}\;\; (i=1,2)$. The resulting Mini Brillouin Zone (MBZ) is shown as a dark hexagon in the center of Fig. ~\ref{fig:BZ}. It should be noticed that the Brillouin zones (BZs) of the two graphene layers are rotated with respect to each other by the same angle as the graphene layer themselves, as shown in Fig. ~\ref{fig:BZ} (blue and orange bigger hexagons). As such, each graphene layer $l=1,2$ has its Dirac points at $\mathbf{K}_{\xi}^{(l)}=-\xi (2 \mathbf{b}_1^{(l)}+\mathbf{b}_2^{(l)})/3$ where $\xi=\pm 1$ labels the valley index. For example, in the same figure we show the $K$ points at the $\xi=-1$ valley for both layers. It turns out that the $K_\xi^{(1)}-K_\xi^{(2)}$ coincides with one of the sides of the MBZ.
\begin{figure}[h]
	\includegraphics*[scale=0.15]{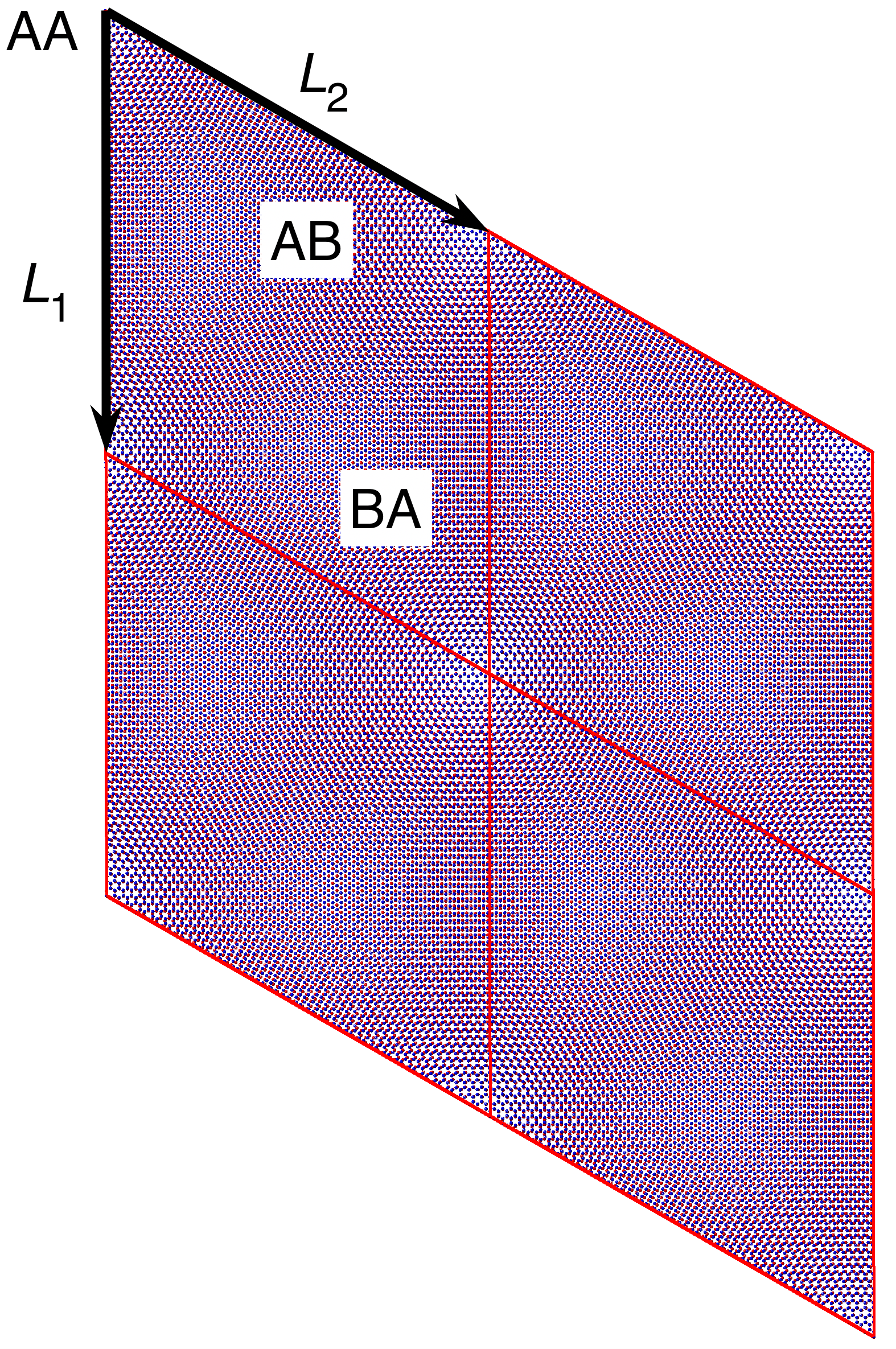}
	\caption{\label{fig:Moire}(Top) The Moir\'e pattern originating from the superposition of two graphene layers after rotation by the commensurate angle 1.08$^\circ$. AA and AB/BA stacking regions
		are highlighted, as well as the supercell sides $\mathbf{L}_1,\mathbf{L}_2$. Blue (red) dots
	represent atoms in the top (bottom) plane.  }
\end{figure}
%
\section{DFT Ab-initio  calculations}
\label{sec:app-DFT}
Density Functional Theory calculations, using the vdW-DF2 exchange-correlation functional \cite{Hamada}, have been carried out using the Vienna Ab initio Simulation Package (VASP) \cite{Kresse1}.  
We used a PAW potential \cite{Blochl,Kresse2} for carbon with the 2p orbitals in valence, and the 1s orbitals frozen in the core. The single particle Bloch waves were expanded
with a plane wave basis set, using a cutoff energy of 400 eV. Sampling of the BZ for the self-consistent (SCF) calculations was 
restricted at the $\Gamma$ point. Single-particle energies at other points in the BZ were obtained by non-SCF calculations. Because of the size 
of the simulation cell, we could only compute one $k$-point at a time, and the reported single-particle energies were therefore referred
to the Fermi energy computed as the energy at the $K$ point. 
The size of the supercell in the direction orthogonal to the layers ($z$-axis) was initially fixed at 10 {\AA}, corresponding to about 
6.5 {\AA} vacuum space,
introduced to prevent periodic replicas of the TBG supercell from interacting with each other. Full relaxation of the atomic positions was carried out
until the residual forces were smaller than 0.002 eV/{\AA}. Additional calculations were repeated using supercells with $z$-axis of 12 {\AA} and 14 {\AA}. A small residual
(maximum) relaxation of less than 0.002{\AA}  was observed as the $z$-axis was increased to 12 {\AA}, but no further relaxation was detectable with the largest 14 {\AA} vacuum space. The initial
relaxation was carried out using 2880 physical cores distributed over 80 nodes of a Cray XC-40 machine, over a period of about 30 days. Calculations with 14 {\AA} vacuum required
5760 cores on 160 nodes to accommodate the extra memory requirements. All symmetries were turned off.
%
\section{Effective Continuum model}
\label{sec:app-ECM}

In the presence of inter-layer interaction the bilayer system can be described by the matrix Hamiltonian 
\begin{equation}
H_\xi(\mathbf{k})=
\left(\begin{array}{cc}
H_\xi^{(1)}(\mathbf{k}) & U_\xi^\dagger\\
U_\xi & H^{(2)}_\xi(\mathbf{k})
\end{array}
\right).
\end{equation}
%


The interlayer Hamiltonian is
\begin{eqnarray}
U_\xi&=&
\left(\begin{array}{cc}
U_{A_2 A_1} & U_{A_2 B_1}\\
U_{B_2 A_1} & U_{B_2 B_1}
\end{array}
\right)=\vspace*{0.1cm}\\
&=& \nonumber
\left(\begin{array}{cc}
u & u'\\
u' & u
\end{array}
\right)+
\left(\begin{array}{cc}
u & u' \omega^{-\xi}\\
u' \omega^{\xi} & u
\end{array}
\right) e^{i \mathbf{G}_1 \cdot \mathbf{r}}+\vspace*{0.1cm}\\
&+& \nonumber
\left(\begin{array}{cc}
u & u' \omega^{\xi}\\
u' \omega^{-\xi} & u
\end{array}
\right) e^{i (\mathbf{G}_1+\mathbf{G}_2) \cdot \mathbf{r}},
\end{eqnarray}
with $\omega= e^{i 2 \pi/3} $. It couples each $\mathbf{k}$ point of the $1^{st}$ layer to a  $\mathbf{k^\prime}$ point of the $2^{nd}$ layer according to the selection rules $\mathbf{k^\prime}=\mathbf{k},\mathbf{k+G_1},\mathbf{k+G_1+G_2}$.
The coefficients $u,\;u'$  are given in Ref.~\onlinecite{Koshino:2018hm}:
\begin{eqnarray}
u(\mathbf{k})&=&-\frac{1}{S_0} \int t \left( \mathbf{R} + d(\mathbf{R}) \mathbf{e}_z \right) e^{-i \mathbf{k}\cdot \mathbf{R}} d^2\mathbf{R} \\
u'(\mathbf{k})&=&-\frac{1}{S_0} \int t\left(\mathbf{R}+d(\mathbf{R}-\boldsymbol{\tau}_1) \mathbf{e}_z\right) e^{-i \mathbf{k}\cdot \mathbf{R}}d^2\mathbf{R}\nonumber
\label{uup}
\end{eqnarray}
where $S_0=\sqrt{3}/2 a^2$ is the unit cell area of the pristine graphene and $t \left(\mathbf{R}\right)$ is the transfer integral between two sites at distance $\mathbf{R}$, originating from the Slater-Koster tight binding parametrization for $p_z$ carbon atoms:
\begin{eqnarray}
t(R)&=&-V_{pp \pi} \left[1-\left(\frac{\mathbf{R}\cdot\hat {\mathbf{z}}}{R}\right)^2\right]-
V_{pp \sigma} \left(\frac{\mathbf{R}\cdot\hat{\mathbf{z}}}{R}\right)^2\:,\\
V_{pp \pi} &=& V_{pp \pi}^0 e^{(R-a_0)/r_0}, \;\;\;\;\; V_{pp \sigma} = V_{pp \sigma}^0 e^{(R-d_0)/r_0}\:\:.
\nonumber
\end{eqnarray}
Here $r_0=0.184 a$ is the decay length of the transfer integral, $a_0=a/\sqrt{3}$ is the first-neighbor distance in graphene, $d_0=0.335$ nm is the intralayer distance, chosen in agreement with that of graphite. $V_{pp \pi}^0=-2.7 eV$ and  $V_{pp \sigma}^0=0.48 eV$ are the in-plane and out of plane nearest-neighbours hopping energy as from Ref.~ \onlinecite{Moon:2013bc}.
We seek for solutions of the kind:
\begin{equation}
\left(\begin{array}{c}
\psi^{A_1}_{n \mathbf{k}}(\mathbf{r})\vspace*{0.1cm}\\
\psi^{B_1}_{n \mathbf{k}}(\mathbf{r})\vspace*{0.1cm}\\
\psi^{A_2}_{n \mathbf{k}}(\mathbf{r})\vspace*{0.1cm}\\
\psi^{B_2}_{n \mathbf{k}}(\mathbf{r})
\end{array}
\right)=  \sum_{\mathbf{G}} 
\left(\begin{array}{c}
c_{n \mathbf{k}}^{A_1} (\mathbf{G})\vspace*{0.1cm} \\
c_{n \mathbf{k}}^{B_1} (\mathbf{G})\vspace*{0.1cm} \\
c_{n \mathbf{k}}^{A_2} (\mathbf{G})\vspace*{0.1cm} \\
c_{n \mathbf{k}}^{B_2} (\mathbf{G}) 
\end{array}
\right)
e^{i (\mathbf{k}+\mathbf{G})\cdot \mathbf{r} }\:.
\end{equation}

The $\mathbf{G}$ point expansion extends, in principle, over the full (infinite) set of $\mathbf{G}$ vectors. However, for numerical purposes this set has to be truncated. We choose a cutoff radius $G_{cut}$, and keep only the $\mathbf{G}$ vectors inside the sphere of radius $G_{cut}$. This is schematically shown in Fig.~\ref{fig:BZ}, where the shadowed circle of radius $G_{cut}$ includes the subset of $\mathbf{G}$ vectors, represented by the red dots. It turns out that the number $N_G$ of vectors to converge the lowest energy states is rather small. The low-energy continuum Hamiltonian matrix has the dimension $D=4 N_{G}$, and  $N_G=19$ (as in the example reported in the figure) allows for a good convergence in an energy shell of few hundreds on meV around the Fermi energy, whereas full convergence, i.e. band energies converged within less than 1 meV, is achieved with $N_G=37$.


\begin{thebibliography}{39}%
\makeatletter
\providecommand \@ifxundefined [1]{%
 \@ifx{#1\undefined}
}%
\providecommand \@ifnum [1]{%
 \ifnum #1\expandafter \@firstoftwo
 \else \expandafter \@secondoftwo
 \fi
}%
\providecommand \@ifx [1]{%
 \ifx #1\expandafter \@firstoftwo
 \else \expandafter \@secondoftwo
 \fi
}%
\providecommand \natexlab [1]{#1}%
\providecommand \enquote  [1]{``#1''}%
\providecommand \bibnamefont  [1]{#1}%
\providecommand \bibfnamefont [1]{#1}%
\providecommand \citenamefont [1]{#1}%
\providecommand \href@noop [0]{\@secondoftwo}%
\providecommand \href [0]{\begingroup \@sanitize@url \@href}%
\providecommand \@href[1]{\@@startlink{#1}\@@href}%
\providecommand \@@href[1]{\endgroup#1\@@endlink}%
\providecommand \@sanitize@url [0]{\catcode `\\12\catcode `\$12\catcode
  `\&12\catcode `\#12\catcode `\^12\catcode `\_12\catcode `\%12\relax}%
\providecommand \@@startlink[1]{}%
\providecommand \@@endlink[0]{}%
\providecommand \url  [0]{\begingroup\@sanitize@url \@url }%
\providecommand \@url [1]{\endgroup\@href {#1}{\urlprefix }}%
\providecommand \urlprefix  [0]{URL }%
\providecommand \Eprint [0]{\href }%
\providecommand \doibase [0]{http://dx.doi.org/}%
\providecommand \selectlanguage [0]{\@gobble}%
\providecommand \bibinfo  [0]{\@secondoftwo}%
\providecommand \bibfield  [0]{\@secondoftwo}%
\providecommand \translation [1]{[#1]}%
\providecommand \BibitemOpen [0]{}%
\providecommand \bibitemStop [0]{}%
\providecommand \bibitemNoStop [0]{.\EOS\space}%
\providecommand \EOS [0]{\spacefactor3000\relax}%
\providecommand \BibitemShut  [1]{\csname bibitem#1\endcsname}%
\let\auto@bib@innerbib\@empty
\bibitem [{\citenamefont {Cao}\ \emph {et~al.}(2018{\natexlab{a}})\citenamefont
  {Cao}, \citenamefont {Fatemi}, \citenamefont {Demir}, \citenamefont {Fang},
  \citenamefont {Tomarken}, \citenamefont {Luo}, \citenamefont
  {Sanchez-Yamagishi}, \citenamefont {Watanabe}, \citenamefont {Taniguchi},
  \citenamefont {Kaxiras}, \citenamefont {Ashoori},\ and\ \citenamefont
  {Jarillo-Herrero}}]{Cao:2018kn}%
  \BibitemOpen
  \bibfield  {author} {\bibinfo {author} {\bibfnamefont {Y.}~\bibnamefont
  {Cao}}, \bibinfo {author} {\bibfnamefont {V.}~\bibnamefont {Fatemi}},
  \bibinfo {author} {\bibfnamefont {A.}~\bibnamefont {Demir}}, \bibinfo
  {author} {\bibfnamefont {S.}~\bibnamefont {Fang}}, \bibinfo {author}
  {\bibfnamefont {S.~L.}\ \bibnamefont {Tomarken}}, \bibinfo {author}
  {\bibfnamefont {J.~Y.}\ \bibnamefont {Luo}}, \bibinfo {author} {\bibfnamefont
  {J.~D.}\ \bibnamefont {Sanchez-Yamagishi}}, \bibinfo {author} {\bibfnamefont
  {K.}~\bibnamefont {Watanabe}}, \bibinfo {author} {\bibfnamefont
  {T.}~\bibnamefont {Taniguchi}}, \bibinfo {author} {\bibfnamefont
  {E.}~\bibnamefont {Kaxiras}}, \bibinfo {author} {\bibfnamefont {R.~C.}\
  \bibnamefont {Ashoori}}, \ and\ \bibinfo {author} {\bibfnamefont
  {P.}~\bibnamefont {Jarillo-Herrero}},\ }\href@noop {} {\bibfield  {journal}
  {\bibinfo  {journal} {Nature Publishing Group}\ }\textbf {\bibinfo {volume}
  {556}},\ \bibinfo {pages} {80} (\bibinfo {year}
  {2018}{\natexlab{a}})}\BibitemShut {NoStop}%
\bibitem [{\citenamefont {Cao}\ \emph {et~al.}(2018{\natexlab{b}})\citenamefont
  {Cao}, \citenamefont {Fatemi}, \citenamefont {Fang}, \citenamefont
  {Watanabe}, \citenamefont {Taniguchi}, \citenamefont {Kaxiras},\ and\
  \citenamefont {Jarillo-Herrero}}]{Cao:2018ff}%
  \BibitemOpen
  \bibfield  {author} {\bibinfo {author} {\bibfnamefont {Y.}~\bibnamefont
  {Cao}}, \bibinfo {author} {\bibfnamefont {V.}~\bibnamefont {Fatemi}},
  \bibinfo {author} {\bibfnamefont {S.}~\bibnamefont {Fang}}, \bibinfo {author}
  {\bibfnamefont {K.}~\bibnamefont {Watanabe}}, \bibinfo {author}
  {\bibfnamefont {T.}~\bibnamefont {Taniguchi}}, \bibinfo {author}
  {\bibfnamefont {E.}~\bibnamefont {Kaxiras}}, \ and\ \bibinfo {author}
  {\bibfnamefont {P.}~\bibnamefont {Jarillo-Herrero}},\ }\href@noop {}
  {\bibfield  {journal} {\bibinfo  {journal} {Nature Publishing Group}\
  }\textbf {\bibinfo {volume} {556}},\ \bibinfo {pages} {43} (\bibinfo {year}
  {2018}{\natexlab{b}})}\BibitemShut {NoStop}%
\bibitem [{\citenamefont {Codecido}\ \emph {et~al.}(2019)\citenamefont
  {Codecido}, \citenamefont {Wang}, \citenamefont {Koester}, \citenamefont
  {Che}, \citenamefont {Tian}, \citenamefont {Lv}, \citenamefont {Tran},
  \citenamefont {Watanabe}, \citenamefont {Taniguchi}, \citenamefont {Zhang},
  \citenamefont {Bockrath},\ and\ \citenamefont {Lau}}]{Codecido:2019wy}%
  \BibitemOpen
  \bibfield  {author} {\bibinfo {author} {\bibfnamefont {E.}~\bibnamefont
  {Codecido}}, \bibinfo {author} {\bibfnamefont {Q.}~\bibnamefont {Wang}},
  \bibinfo {author} {\bibfnamefont {R.}~\bibnamefont {Koester}}, \bibinfo
  {author} {\bibfnamefont {S.}~\bibnamefont {Che}}, \bibinfo {author}
  {\bibfnamefont {H.}~\bibnamefont {Tian}}, \bibinfo {author} {\bibfnamefont
  {R.}~\bibnamefont {Lv}}, \bibinfo {author} {\bibfnamefont {S.}~\bibnamefont
  {Tran}}, \bibinfo {author} {\bibfnamefont {K.}~\bibnamefont {Watanabe}},
  \bibinfo {author} {\bibfnamefont {T.}~\bibnamefont {Taniguchi}}, \bibinfo
  {author} {\bibfnamefont {F.}~\bibnamefont {Zhang}}, \bibinfo {author}
  {\bibfnamefont {M.}~\bibnamefont {Bockrath}}, \ and\ \bibinfo {author}
  {\bibfnamefont {C.~N.}\ \bibnamefont {Lau}},\ }\href@noop {} {\  (\bibinfo
  {year} {2019})},\ \Eprint {http://arxiv.org/abs/arXiv:1902.05151
  [cond-mat.mes-hall]} {arXiv:1902.05151 [cond-mat.mes-hall]} \BibitemShut
  {NoStop}%
\bibitem [{\citenamefont {Choi}\ and\ \citenamefont
  {Choi}(2018)}]{Choi:2018cu}%
  \BibitemOpen
  \bibfield  {author} {\bibinfo {author} {\bibfnamefont {Y.~W.}\ \bibnamefont
  {Choi}}\ and\ \bibinfo {author} {\bibfnamefont {H.~J.}\ \bibnamefont
  {Choi}},\ }\href {\doibase 10.1103/PhysRevB.98.241412} {\bibfield  {journal}
  {\bibinfo  {journal} {Phys. Rev. B}\ }\textbf {\bibinfo {volume} {98}},\
  \bibinfo {pages} {241412(R)} (\bibinfo {year} {2018})}\BibitemShut {NoStop}%
\bibitem [{\citenamefont {Sboychakov}\ \emph {et~al.}(2018)\citenamefont
  {Sboychakov}, \citenamefont {Rozhkov}, \citenamefont {Rakhmanov},\ and\
  \citenamefont {Nori}}]{Sboychakov:2018tp}%
  \BibitemOpen
  \bibfield  {author} {\bibinfo {author} {\bibfnamefont {A.~O.}\ \bibnamefont
  {Sboychakov}}, \bibinfo {author} {\bibfnamefont {A.~V.}\ \bibnamefont
  {Rozhkov}}, \bibinfo {author} {\bibfnamefont {A.~L.}\ \bibnamefont
  {Rakhmanov}}, \ and\ \bibinfo {author} {\bibfnamefont {F.}~\bibnamefont
  {Nori}},\ }\href@noop {} {\bibfield  {journal} {\bibinfo  {journal} {arXiv}\
  } (\bibinfo {year} {2018})},\ \Eprint
  {http://arxiv.org/abs/arxiv:1807.08190v1} {arxiv:1807.08190v1} \BibitemShut
  {NoStop}%
\bibitem [{\citenamefont {Ribeiro-Palau}\ \emph {et~al.}(2018)\citenamefont
  {Ribeiro-Palau}, \citenamefont {Zhang}, \citenamefont {Watanabe},
  \citenamefont {Taniguchi}, \citenamefont {Hone},\ and\ \citenamefont
  {Dean}}]{RibeiroPalau:2018ki}%
  \BibitemOpen
  \bibfield  {author} {\bibinfo {author} {\bibfnamefont {R.}~\bibnamefont
  {Ribeiro-Palau}}, \bibinfo {author} {\bibfnamefont {C.}~\bibnamefont
  {Zhang}}, \bibinfo {author} {\bibfnamefont {K.}~\bibnamefont {Watanabe}},
  \bibinfo {author} {\bibfnamefont {T.}~\bibnamefont {Taniguchi}}, \bibinfo
  {author} {\bibfnamefont {J.}~\bibnamefont {Hone}}, \ and\ \bibinfo {author}
  {\bibfnamefont {C.~R.}\ \bibnamefont {Dean}},\ }\href@noop {} {\bibfield
  {journal} {\bibinfo  {journal} {Science}\ }\textbf {\bibinfo {volume}
  {361}},\ \bibinfo {pages} {690} (\bibinfo {year} {2018})}\BibitemShut
  {NoStop}%
\bibitem [{\citenamefont {Geim}\ and\ \citenamefont
  {Grigorieva}(2014)}]{Geim:2014hf}%
  \BibitemOpen
  \bibfield  {author} {\bibinfo {author} {\bibfnamefont {A.~K.}\ \bibnamefont
  {Geim}}\ and\ \bibinfo {author} {\bibfnamefont {I.~V.}\ \bibnamefont
  {Grigorieva}},\ }\href@noop {} {\bibfield  {journal} {\bibinfo  {journal}
  {Nature}\ }\textbf {\bibinfo {volume} {499}},\ \bibinfo {pages} {419}
  (\bibinfo {year} {2014})}\BibitemShut {NoStop}%
\bibitem [{\citenamefont {Cantele}\ and\ \citenamefont
  {Ninno}(2017)}]{PhysRevMaterials.1.014002}%
  \BibitemOpen
  \bibfield  {author} {\bibinfo {author} {\bibfnamefont {G.}~\bibnamefont
  {Cantele}}\ and\ \bibinfo {author} {\bibfnamefont {D.}~\bibnamefont
  {Ninno}},\ }\href {\doibase 10.1103/PhysRevMaterials.1.014002} {\bibfield
  {journal} {\bibinfo  {journal} {Phys. Rev. Materials}\ }\textbf {\bibinfo
  {volume} {1}},\ \bibinfo {pages} {014002} (\bibinfo {year}
  {2017})}\BibitemShut {NoStop}%
\bibitem [{\citenamefont {Borriello}\ \emph {et~al.}(2012)\citenamefont
  {Borriello}, \citenamefont {Cantele},\ and\ \citenamefont
  {Ninno}}]{Borriello:2012ja}%
  \BibitemOpen
  \bibfield  {author} {\bibinfo {author} {\bibfnamefont {I.}~\bibnamefont
  {Borriello}}, \bibinfo {author} {\bibfnamefont {G.}~\bibnamefont {Cantele}},
  \ and\ \bibinfo {author} {\bibfnamefont {D.}~\bibnamefont {Ninno}},\
  }\href@noop {} {\bibfield  {journal} {\bibinfo  {journal} {Nanoscale}\
  }\textbf {\bibinfo {volume} {5}},\ \bibinfo {pages} {291} (\bibinfo {year}
  {2012})}\BibitemShut {NoStop}%
\bibitem [{\citenamefont {Cantele}\ \emph {et~al.}(2009)\citenamefont
  {Cantele}, \citenamefont {Lee}, \citenamefont {Ninno},\ and\ \citenamefont
  {Marzari}}]{Cantele:2009de}%
  \BibitemOpen
  \bibfield  {author} {\bibinfo {author} {\bibfnamefont {G.}~\bibnamefont
  {Cantele}}, \bibinfo {author} {\bibfnamefont {Y.-S.}\ \bibnamefont {Lee}},
  \bibinfo {author} {\bibfnamefont {D.}~\bibnamefont {Ninno}}, \ and\ \bibinfo
  {author} {\bibfnamefont {N.}~\bibnamefont {Marzari}},\ }\href@noop {}
  {\bibfield  {journal} {\bibinfo  {journal} {Nano Lett}\ }\textbf {\bibinfo
  {volume} {9}},\ \bibinfo {pages} {3425} (\bibinfo {year} {2009})}\BibitemShut
  {NoStop}%
\bibitem [{\citenamefont {Song}\ \emph {et~al.}(2018)\citenamefont {Song},
  \citenamefont {Wang}, \citenamefont {Shi}, \citenamefont {Li}, \citenamefont
  {Fang},\ and\ \citenamefont {Bernevig}}]{Song:2018ul}%
  \BibitemOpen
  \bibfield  {author} {\bibinfo {author} {\bibfnamefont {Z.}~\bibnamefont
  {Song}}, \bibinfo {author} {\bibfnamefont {Z.}~\bibnamefont {Wang}}, \bibinfo
  {author} {\bibfnamefont {W.}~\bibnamefont {Shi}}, \bibinfo {author}
  {\bibfnamefont {G.}~\bibnamefont {Li}}, \bibinfo {author} {\bibfnamefont
  {C.}~\bibnamefont {Fang}}, \ and\ \bibinfo {author} {\bibfnamefont {B.~A.}\
  \bibnamefont {Bernevig}},\ }\href@noop {} {\  (\bibinfo {year} {2018})},\
  \Eprint {http://arxiv.org/abs/arxiv:1807.10676} {arxiv:1807.10676}
  \BibitemShut {NoStop}%
\bibitem [{\citenamefont {Hejazi}\ \emph {et~al.}(2019)\citenamefont {Hejazi},
  \citenamefont {Liu}, \citenamefont {Shapourian}, \citenamefont {Chen},\ and\
  \citenamefont {Balents}}]{Hejazi:2019dz}%
  \BibitemOpen
  \bibfield  {author} {\bibinfo {author} {\bibfnamefont {K.}~\bibnamefont
  {Hejazi}}, \bibinfo {author} {\bibfnamefont {C.}~\bibnamefont {Liu}},
  \bibinfo {author} {\bibfnamefont {H.}~\bibnamefont {Shapourian}}, \bibinfo
  {author} {\bibfnamefont {X.}~\bibnamefont {Chen}}, \ and\ \bibinfo {author}
  {\bibfnamefont {L.}~\bibnamefont {Balents}},\ }\href@noop {} {\bibfield
  {journal} {\bibinfo  {journal} {Phys. Rev. B}\ }\textbf {\bibinfo {volume}
  {99}},\ \bibinfo {pages} {035111} (\bibinfo {year} {2019})}\BibitemShut
  {NoStop}%
\bibitem [{\citenamefont {Liu}\ \emph {et~al.}(2018)\citenamefont {Liu},
  \citenamefont {Liu},\ and\ \citenamefont {Dai}}]{Liu:2018}%
  \BibitemOpen
  \bibfield  {author} {\bibinfo {author} {\bibfnamefont {J.}~\bibnamefont
  {Liu}}, \bibinfo {author} {\bibfnamefont {J.}~\bibnamefont {Liu}}, \ and\
  \bibinfo {author} {\bibfnamefont {X.}~\bibnamefont {Dai}},\ }\href@noop {} {\
   (\bibinfo {year} {2018})},\ \Eprint {http://arxiv.org/abs/arXiv:1810.03103
  [cond-mat.mes-hall]} {arXiv:1810.03103 [cond-mat.mes-hall]} \BibitemShut
  {NoStop}%
\bibitem [{\citenamefont {Hasan}\ and\ \citenamefont
  {Kane}(2010)}]{RevModPhys.82.3045}%
  \BibitemOpen
  \bibfield  {author} {\bibinfo {author} {\bibfnamefont {M.~Z.}\ \bibnamefont
  {Hasan}}\ and\ \bibinfo {author} {\bibfnamefont {C.~L.}\ \bibnamefont
  {Kane}},\ }\href {\doibase 10.1103/RevModPhys.82.3045} {\bibfield  {journal}
  {\bibinfo  {journal} {Rev. Mod. Phys.}\ }\textbf {\bibinfo {volume} {82}},\
  \bibinfo {pages} {3045} (\bibinfo {year} {2010})}\BibitemShut {NoStop}%
\bibitem [{\citenamefont {Bercioux}\ and\ \citenamefont
  {Lucignano}(2015)}]{0034-4885-78-10-106001}%
  \BibitemOpen
  \bibfield  {author} {\bibinfo {author} {\bibfnamefont {D.}~\bibnamefont
  {Bercioux}}\ and\ \bibinfo {author} {\bibfnamefont {P.}~\bibnamefont
  {Lucignano}},\ }\href {http://stacks.iop.org/0034-4885/78/i=10/a=106001}
  {\bibfield  {journal} {\bibinfo  {journal} {Reports on Progress in Physics}\
  }\textbf {\bibinfo {volume} {78}},\ \bibinfo {pages} {106001} (\bibinfo
  {year} {2015})}\BibitemShut {NoStop}%
\bibitem [{\citenamefont {Lucignano}\ \emph {et~al.}(2008)\citenamefont
  {Lucignano}, \citenamefont {Raimondi},\ and\ \citenamefont
  {Tagliacozzo}}]{PhysRevB.78.035336}%
  \BibitemOpen
  \bibfield  {author} {\bibinfo {author} {\bibfnamefont {P.}~\bibnamefont
  {Lucignano}}, \bibinfo {author} {\bibfnamefont {R.}~\bibnamefont {Raimondi}},
  \ and\ \bibinfo {author} {\bibfnamefont {A.}~\bibnamefont {Tagliacozzo}},\
  }\href {\doibase 10.1103/PhysRevB.78.035336} {\bibfield  {journal} {\bibinfo
  {journal} {Phys. Rev. B}\ }\textbf {\bibinfo {volume} {78}},\ \bibinfo
  {pages} {035336} (\bibinfo {year} {2008})}\BibitemShut {NoStop}%
\bibitem [{\citenamefont {Lopes~dos Santos}\ \emph {et~al.}(2007)\citenamefont
  {Lopes~dos Santos}, \citenamefont {Peres},\ and\ \citenamefont
  {Castro~Neto}}]{LopesdosSantos:2007fg}%
  \BibitemOpen
  \bibfield  {author} {\bibinfo {author} {\bibfnamefont {J.~M.~B.}\
  \bibnamefont {Lopes~dos Santos}}, \bibinfo {author} {\bibfnamefont
  {N.~M.~R.}\ \bibnamefont {Peres}}, \ and\ \bibinfo {author} {\bibfnamefont
  {A.~H.}\ \bibnamefont {Castro~Neto}},\ }\href@noop {} {\bibfield  {journal}
  {\bibinfo  {journal} {Phys Rev Lett}\ }\textbf {\bibinfo {volume} {99}},\
  \bibinfo {pages} {256802} (\bibinfo {year} {2007})}\BibitemShut {NoStop}%
\bibitem [{\citenamefont {Lopes~dos Santos}\ \emph {et~al.}(2012)\citenamefont
  {Lopes~dos Santos}, \citenamefont {Peres},\ and\ \citenamefont
  {Castro~Neto}}]{LopesdosSantos:2007fga}%
  \BibitemOpen
  \bibfield  {author} {\bibinfo {author} {\bibfnamefont {J.~M.~B.}\
  \bibnamefont {Lopes~dos Santos}}, \bibinfo {author} {\bibfnamefont
  {N.~M.~R.}\ \bibnamefont {Peres}}, \ and\ \bibinfo {author} {\bibfnamefont
  {A.~H.}\ \bibnamefont {Castro~Neto}},\ }\href@noop {} {\bibfield  {journal}
  {\bibinfo  {journal} {Phys. Rev. B}\ }\textbf {\bibinfo {volume} {86}},\
  \bibinfo {pages} {155449} (\bibinfo {year} {2012})}\BibitemShut {NoStop}%
\bibitem [{\citenamefont {Mele}(2010)}]{Mele:2010jg}%
  \BibitemOpen
  \bibfield  {author} {\bibinfo {author} {\bibfnamefont {E.~J.}\ \bibnamefont
  {Mele}},\ }\href@noop {} {\bibfield  {journal} {\bibinfo  {journal} {Phys.
  Rev. B}\ }\textbf {\bibinfo {volume} {81}},\ \bibinfo {pages} {161405(R)}
  (\bibinfo {year} {2010})}\BibitemShut {NoStop}%
\bibitem [{\citenamefont {Bistritzer}\ and\ \citenamefont
  {MacDonald}(2011)}]{Bistritzer:2011ho}%
  \BibitemOpen
  \bibfield  {author} {\bibinfo {author} {\bibfnamefont {R.}~\bibnamefont
  {Bistritzer}}\ and\ \bibinfo {author} {\bibfnamefont {A.~H.}\ \bibnamefont
  {MacDonald}},\ }\href@noop {} {\bibfield  {journal} {\bibinfo  {journal}
  {Proceedings of the National Academy of Sciences}\ }\textbf {\bibinfo
  {volume} {108}},\ \bibinfo {pages} {12233} (\bibinfo {year}
  {2011})}\BibitemShut {NoStop}%
\bibitem [{\citenamefont {Moon}\ and\ \citenamefont
  {Koshino}(2013)}]{Moon:2013bc}%
  \BibitemOpen
  \bibfield  {author} {\bibinfo {author} {\bibfnamefont {P.}~\bibnamefont
  {Moon}}\ and\ \bibinfo {author} {\bibfnamefont {M.}~\bibnamefont {Koshino}},\
  }\href@noop {} {\bibfield  {journal} {\bibinfo  {journal} {Phys. Rev. B}\
  }\textbf {\bibinfo {volume} {87}},\ \bibinfo {pages} {205404} (\bibinfo
  {year} {2013})}\BibitemShut {NoStop}%
\bibitem [{\citenamefont {Koshino}\ \emph {et~al.}(2018)\citenamefont
  {Koshino}, \citenamefont {Yuan}, \citenamefont {Koretsune}, \citenamefont
  {Ochi}, \citenamefont {Kuroki},\ and\ \citenamefont {Fu}}]{Koshino:2018hm}%
  \BibitemOpen
  \bibfield  {author} {\bibinfo {author} {\bibfnamefont {M.}~\bibnamefont
  {Koshino}}, \bibinfo {author} {\bibfnamefont {N.~F.~Q.}\ \bibnamefont
  {Yuan}}, \bibinfo {author} {\bibfnamefont {T.}~\bibnamefont {Koretsune}},
  \bibinfo {author} {\bibfnamefont {M.}~\bibnamefont {Ochi}}, \bibinfo {author}
  {\bibfnamefont {K.}~\bibnamefont {Kuroki}}, \ and\ \bibinfo {author}
  {\bibfnamefont {L.}~\bibnamefont {Fu}},\ }\href@noop {} {\bibfield  {journal}
  {\bibinfo  {journal} {Phys. Rev. X}\ }\textbf {\bibinfo {volume} {8}},\
  \bibinfo {pages} {031087} (\bibinfo {year} {2018})}\BibitemShut {NoStop}%
\bibitem [{\citenamefont {Jung}\ and\ \citenamefont
  {MacDonald}(2013)}]{Jung:2013ix}%
  \BibitemOpen
  \bibfield  {author} {\bibinfo {author} {\bibfnamefont {J.}~\bibnamefont
  {Jung}}\ and\ \bibinfo {author} {\bibfnamefont {A.~H.}\ \bibnamefont
  {MacDonald}},\ }\href@noop {} {\bibfield  {journal} {\bibinfo  {journal}
  {Phys. Rev. B}\ }\textbf {\bibinfo {volume} {87}},\ \bibinfo {pages} {195450}
  (\bibinfo {year} {2013})}\BibitemShut {NoStop}%
\bibitem [{\citenamefont {Sboychakov}\ \emph {et~al.}(2015)\citenamefont
  {Sboychakov}, \citenamefont {Rakhmanov}, \citenamefont {Rozhkov},\ and\
  \citenamefont {Nori}}]{Sboychakov:2015jx}%
  \BibitemOpen
  \bibfield  {author} {\bibinfo {author} {\bibfnamefont {A.~O.}\ \bibnamefont
  {Sboychakov}}, \bibinfo {author} {\bibfnamefont {A.~L.}\ \bibnamefont
  {Rakhmanov}}, \bibinfo {author} {\bibfnamefont {A.~V.}\ \bibnamefont
  {Rozhkov}}, \ and\ \bibinfo {author} {\bibfnamefont {F.}~\bibnamefont
  {Nori}},\ }\href@noop {} {\bibfield  {journal} {\bibinfo  {journal} {Phys.
  Rev. B}\ }\textbf {\bibinfo {volume} {92}},\ \bibinfo {pages} {075402}
  (\bibinfo {year} {2015})}\BibitemShut {NoStop}%
\bibitem [{\citenamefont {Fang}\ and\ \citenamefont
  {Kaxiras}(2016)}]{Fang:2016iq}%
  \BibitemOpen
  \bibfield  {author} {\bibinfo {author} {\bibfnamefont {S.}~\bibnamefont
  {Fang}}\ and\ \bibinfo {author} {\bibfnamefont {E.}~\bibnamefont {Kaxiras}},\
  }\href@noop {} {\bibfield  {journal} {\bibinfo  {journal} {Phys. Rev. B}\
  }\textbf {\bibinfo {volume} {93}},\ \bibinfo {pages} {235153} (\bibinfo
  {year} {2016})}\BibitemShut {NoStop}%
\bibitem [{\citenamefont {Lin}\ and\ \citenamefont
  {Tom\'anek}(2018)}]{Lin:2018fy}%
  \BibitemOpen
  \bibfield  {author} {\bibinfo {author} {\bibfnamefont {X.}~\bibnamefont
  {Lin}}\ and\ \bibinfo {author} {\bibfnamefont {D.}~\bibnamefont
  {Tom\'anek}},\ }\href {\doibase 10.1103/PhysRevB.98.081410} {\bibfield
  {journal} {\bibinfo  {journal} {Phys. Rev. B}\ }\textbf {\bibinfo {volume}
  {98}},\ \bibinfo {pages} {081410(R)} (\bibinfo {year} {2018})}\BibitemShut
  {NoStop}%
\bibitem [{\citenamefont {Kang}\ and\ \citenamefont
  {Vafek}(2018)}]{Kang:2018et}%
  \BibitemOpen
  \bibfield  {author} {\bibinfo {author} {\bibfnamefont {J.}~\bibnamefont
  {Kang}}\ and\ \bibinfo {author} {\bibfnamefont {O.}~\bibnamefont {Vafek}},\
  }\href@noop {} {\bibfield  {journal} {\bibinfo  {journal} {Phys. Rev. X}\
  }\textbf {\bibinfo {volume} {8}},\ \bibinfo {pages} {031088} (\bibinfo {year}
  {2018})}\BibitemShut {NoStop}%
\bibitem [{\citenamefont {Angeli}\ \emph {et~al.}(2018)\citenamefont {Angeli},
  \citenamefont {Mandelli}, \citenamefont {Valli}, \citenamefont {Amaricci},
  \citenamefont {Capone}, \citenamefont {Tosatti},\ and\ \citenamefont
  {Fabrizio}}]{Angeli:2018wy}%
  \BibitemOpen
  \bibfield  {author} {\bibinfo {author} {\bibfnamefont {M.}~\bibnamefont
  {Angeli}}, \bibinfo {author} {\bibfnamefont {D.}~\bibnamefont {Mandelli}},
  \bibinfo {author} {\bibfnamefont {A.}~\bibnamefont {Valli}}, \bibinfo
  {author} {\bibfnamefont {A.}~\bibnamefont {Amaricci}}, \bibinfo {author}
  {\bibfnamefont {M.}~\bibnamefont {Capone}}, \bibinfo {author} {\bibfnamefont
  {E.}~\bibnamefont {Tosatti}}, \ and\ \bibinfo {author} {\bibfnamefont
  {M.}~\bibnamefont {Fabrizio}},\ }\href@noop {} {\  (\bibinfo {year}
  {2018})},\ \Eprint {http://arxiv.org/abs/arxiv:1809.11140v1}
  {arxiv:1809.11140v1} \BibitemShut {NoStop}%
\bibitem [{\citenamefont {Nam~Do}\ \emph {et~al.}(2018)\citenamefont {Nam~Do},
  \citenamefont {Anh~Le},\ and\ \citenamefont {Bercioux}}]{Bercioux:2018}%
  \BibitemOpen
  \bibfield  {author} {\bibinfo {author} {\bibfnamefont {V.}~\bibnamefont
  {Nam~Do}}, \bibinfo {author} {\bibfnamefont {H.}~\bibnamefont {Anh~Le}}, \
  and\ \bibinfo {author} {\bibfnamefont {D.}~\bibnamefont {Bercioux}},\
  }\href@noop {} {\  (\bibinfo {year} {2018})},\ \Eprint
  {http://arxiv.org/abs/arXiv:1901.02794} {arXiv:1901.02794} \BibitemShut
  {NoStop}%
\bibitem [{\citenamefont {Le}\ and\ \citenamefont
  {Do}(2018)}]{PhysRevB.97.125136}%
  \BibitemOpen
  \bibfield  {author} {\bibinfo {author} {\bibfnamefont {H.~A.}\ \bibnamefont
  {Le}}\ and\ \bibinfo {author} {\bibfnamefont {V.~N.}\ \bibnamefont {Do}},\
  }\href {\doibase 10.1103/PhysRevB.97.125136} {\bibfield  {journal} {\bibinfo
  {journal} {Phys. Rev. B}\ }\textbf {\bibinfo {volume} {97}},\ \bibinfo
  {pages} {125136} (\bibinfo {year} {2018})}\BibitemShut {NoStop}%
\bibitem [{\citenamefont {Uchida}\ \emph {et~al.}(2014)\citenamefont {Uchida},
  \citenamefont {Furuya}, \citenamefont {Iwata},\ and\ \citenamefont
  {Oshiyama}}]{Uchida:2014}%
  \BibitemOpen
  \bibfield  {author} {\bibinfo {author} {\bibfnamefont {K.}~\bibnamefont
  {Uchida}}, \bibinfo {author} {\bibfnamefont {S.}~\bibnamefont {Furuya}},
  \bibinfo {author} {\bibfnamefont {J.-I.}\ \bibnamefont {Iwata}}, \ and\
  \bibinfo {author} {\bibfnamefont {A.}~\bibnamefont {Oshiyama}},\ }\href@noop
  {} {\bibfield  {journal} {\bibinfo  {journal} {Phys. Rev. B}\ }\textbf
  {\bibinfo {volume} {90}},\ \bibinfo {pages} {155451} (\bibinfo {year}
  {2014})}\BibitemShut {NoStop}%
\bibitem [{\citenamefont {van Wijk}\ \emph {et~al.}(2015)\citenamefont {van
  Wijk}, \citenamefont {Schuring}, \citenamefont {Katsnelson},\ and\
  \citenamefont {Fasolino}}]{Fasolino:2015}%
  \BibitemOpen
  \bibfield  {author} {\bibinfo {author} {\bibfnamefont {M.~M.}\ \bibnamefont
  {van Wijk}}, \bibinfo {author} {\bibfnamefont {A.}~\bibnamefont {Schuring}},
  \bibinfo {author} {\bibfnamefont {M.~I.}\ \bibnamefont {Katsnelson}}, \ and\
  \bibinfo {author} {\bibfnamefont {A.}~\bibnamefont {Fasolino}},\ }\href@noop
  {} {\bibfield  {journal} {\bibinfo  {journal} {2D Materials}\ }\textbf
  {\bibinfo {volume} {2}},\ \bibinfo {pages} {034010} (\bibinfo {year}
  {2015})}\BibitemShut {NoStop}%
\bibitem [{\citenamefont {Lin}\ \emph {et~al.}(2018)\citenamefont {Lin},
  \citenamefont {Liu},\ and\ \citenamefont {Tom\'anek}}]{Tomanek:2018}%
  \BibitemOpen
  \bibfield  {author} {\bibinfo {author} {\bibfnamefont {X.}~\bibnamefont
  {Lin}}, \bibinfo {author} {\bibfnamefont {D.}~\bibnamefont {Liu}}, \ and\
  \bibinfo {author} {\bibfnamefont {D.}~\bibnamefont {Tom\'anek}},\ }\href@noop
  {} {\bibfield  {journal} {\bibinfo  {journal} {Phys. Rev. B}\ }\textbf
  {\bibinfo {volume} {98}},\ \bibinfo {pages} {195432} (\bibinfo {year}
  {2018})}\BibitemShut {NoStop}%
\bibitem [{\citenamefont {Gargiulo}\ and\ \citenamefont
  {Yazyev}(2018)}]{Gargiulo:2018bj}%
  \BibitemOpen
  \bibfield  {author} {\bibinfo {author} {\bibfnamefont {F.}~\bibnamefont
  {Gargiulo}}\ and\ \bibinfo {author} {\bibfnamefont {O.~V.}\ \bibnamefont
  {Yazyev}},\ }\href@noop {} {\bibfield  {journal} {\bibinfo  {journal} {2D
  Mater.}\ }\textbf {\bibinfo {volume} {5}},\ \bibinfo {pages} {015019}
  (\bibinfo {year} {2018})}\BibitemShut {NoStop}%
\bibitem [{\citenamefont {Jain}\ \emph {et~al.}(2016)\citenamefont {Jain},
  \citenamefont {Juri{\v{c}}i{\'{c}}},\ and\ \citenamefont
  {Barkema}}]{Jain_2016}%
  \BibitemOpen
  \bibfield  {author} {\bibinfo {author} {\bibfnamefont {S.~K.}\ \bibnamefont
  {Jain}}, \bibinfo {author} {\bibfnamefont {V.}~\bibnamefont
  {Juri{\v{c}}i{\'{c}}}}, \ and\ \bibinfo {author} {\bibfnamefont {G.~T.}\
  \bibnamefont {Barkema}},\ }\href {\doibase 10.1088/2053-1583/4/1/015018}
  {\bibfield  {journal} {\bibinfo  {journal} {2D Materials}\ }\textbf {\bibinfo
  {volume} {4}},\ \bibinfo {pages} {015018} (\bibinfo {year}
  {2016})}\BibitemShut {NoStop}%
\bibitem [{\citenamefont {Hamada}(2014)}]{Hamada}%
  \BibitemOpen
  \bibfield  {author} {\bibinfo {author} {\bibfnamefont {I.}~\bibnamefont
  {Hamada}},\ }\href {\doibase 10.1103/PhysRevB.89.121103} {\bibfield
  {journal} {\bibinfo  {journal} {Phys. Rev. B}\ }\textbf {\bibinfo {volume}
  {89}},\ \bibinfo {pages} {121103(R)} (\bibinfo {year} {2014})}\BibitemShut
  {NoStop}%
\bibitem [{\citenamefont {Kresse}\ and\ \citenamefont
  {Furthm\"uller}(1996)}]{Kresse1}%
  \BibitemOpen
  \bibfield  {author} {\bibinfo {author} {\bibfnamefont {G.}~\bibnamefont
  {Kresse}}\ and\ \bibinfo {author} {\bibfnamefont {J.}~\bibnamefont
  {Furthm\"uller}},\ }\href {\doibase 10.1103/PhysRevB.54.11169} {\bibfield
  {journal} {\bibinfo  {journal} {Phys. Rev. B}\ }\textbf {\bibinfo {volume}
  {54}},\ \bibinfo {pages} {11169} (\bibinfo {year} {1996})}\BibitemShut
  {NoStop}%
\bibitem [{\citenamefont {Bl\"ochl}(1994)}]{Blochl}%
  \BibitemOpen
  \bibfield  {author} {\bibinfo {author} {\bibfnamefont {P.~E.}\ \bibnamefont
  {Bl\"ochl}},\ }\href {\doibase 10.1103/PhysRevB.50.17953} {\bibfield
  {journal} {\bibinfo  {journal} {Phys. Rev. B}\ }\textbf {\bibinfo {volume}
  {50}},\ \bibinfo {pages} {17953} (\bibinfo {year} {1994})}\BibitemShut
  {NoStop}%
\bibitem [{\citenamefont {Kresse}\ and\ \citenamefont
  {Joubert}(1999)}]{Kresse2}%
  \BibitemOpen
  \bibfield  {author} {\bibinfo {author} {\bibfnamefont {G.}~\bibnamefont
  {Kresse}}\ and\ \bibinfo {author} {\bibfnamefont {D.}~\bibnamefont
  {Joubert}},\ }\href {\doibase 10.1103/PhysRevB.59.1758} {\bibfield  {journal}
  {\bibinfo  {journal} {Phys. Rev. B}\ }\textbf {\bibinfo {volume} {59}},\
  \bibinfo {pages} {1758} (\bibinfo {year} {1999})}\BibitemShut {NoStop}%
\end{thebibliography}
%

\end{document}